\documentclass[final,3p,times,twocolumn]{elsarticle}

\usepackage{amssymb}
\usepackage{amsmath,amsfonts}
\usepackage{algorithmic}
\usepackage{algorithm}
\usepackage{array}
\usepackage{textcomp}
\usepackage{stfloats}
\usepackage{url}
\usepackage{verbatim}
\usepackage{graphicx}
\usepackage{subfigure}
\usepackage{multirow}
\usepackage{multicol}
\usepackage[switch]{lineno}
\usepackage{color}

\usepackage{booktabs}
\biboptions{numbers,sort&compress}
\usepackage[colorlinks,linkcolor=blue]{hyperref}

\begin{document}

\begin{frontmatter}


\title{Causal SAR ATR with Limited Data via Dual Invariance}


\author[1]{Chenwei Wang}
\author[2]{You Qin}
\author[2]{Li Li}
\author[1]{Siyi Luo}
\author[1]{Yulin Huang\corref{cor1}}
\author[1]{Jifang Pei}
\author[1]{Yin Zhang}
\author[1]{Jianyu Yang}
\cortext[cor1]{Corresponding author}

\affiliation[1]{organization={School of Information and Communication Engineering, University of Electronic Science and Technology of China}, city={Chengdu}, postcode={611731}, country={China}}
\affiliation[2]{organization={School of Computing, National University of Singapore}, city={Singapore}, postcode={119245}, country={Singapore}}


\begin{abstract}
Synthetic aperture radar automatic target recognition (SAR ATR) with limited data has recently been a hot research topic to enhance weak generalization. 
Despite many excellent methods being proposed, a fundamental theory is lacked to explain what problem the limited SAR data causes, leading to weak generalization of ATR. 
In this paper, we establish a causal ATR model demonstrating that noise $N$ that could be blocked with ample SAR data, becomes a confounder with limited data for recognition. 
As a result, it has a detrimental causal effect damaging the efficacy of feature $X$ extracted from SAR images, leading to weak generalization of SAR ATR with limited data.
The effect of $N$ on feature can be estimated and eliminated by using backdoor adjustment to pursue the direct causality between $X$ and the predicted class $Y$.
However, it is difficult for SAR images to precisely estimate and eliminated the effect of $N$ on $X$. 
The presence of various interference types in SAR images, such as cluster and speckle, and the inter-class similarity of SAR images, which can be mistaken for the effect of $N$, complicates the precise estimation of $N$'s effect.
The limited SAR data scarcely powers the majority of existing optimization losses based on empirical risk minimization (ERM), thus making it difficult to effectively eliminate $N$'s effect. 
To tackle with difficult estimation and elimination of $N$'s effect, we propose a dual invariance comprising the inner-class invariant proxy and the noise-invariance loss. 
Motivated by tackling change with invariance, the inner-class invariant proxy facilitates precise estimation of $N$'s effect on $X$ by obtaining accurate invariant features for each class with the limited data. 
The noise-invariance loss transitions the ERM's data quantity necessity into a need for noise environment annotations, effectively eliminating $N$'s effect on $X$ by cleverly applying the previous $N$'s estimation as the noise environment annotations. 
Finally, the proposed causal ATR via dual invariance not only unravels the key problem caused by limited data, but also derives an effective principled solution.
Experiments on three benchmark datasets indicate that the proposed method achieves superior performance. 
The soundness and effectiveness of the proposed method are further demonstrated through comprehensive ablation experiments. We will release our codes and more experimental results at 
\url{https://github.com/cwwangSARATR/SARATR_Causal_Dual_Invariance}.

\end{abstract}

\begin{graphicalabstract}
\centering
\includegraphics[width=0.95\textwidth]{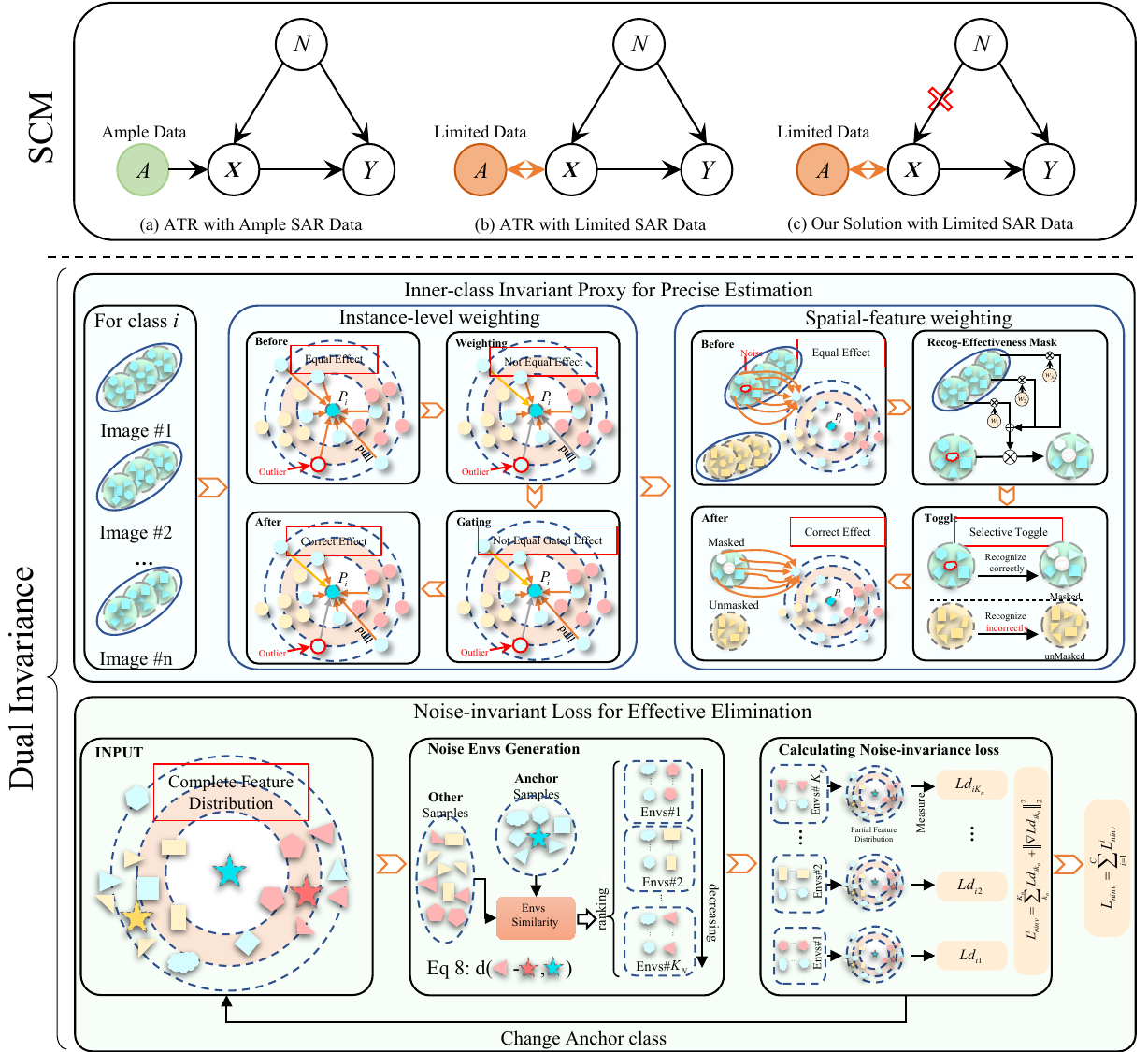}
\end{graphicalabstract}


\end{frontmatter}


\section{Introduction}
\label{introduction}
Synthetic aperture radar (SAR) is a versatile remote sensing technology used in various civilian and military applications, providing high-resolution images regardless of time or weather conditions \cite{intro1}. 
Automatic target recognition (ATR) is a critical SAR application and has been developed for half a century \cite{intro2,addnew1, reff1, reff2, reff3, my1,addnew2,addnew3,addnew4,addnew5,addnew6}.
In recent ten years, ATR 's researches have achieved significant recognition performance under the promotion of deep learning technology \cite{ATR3,isprs3,ATR1,isprs2,isprs6,addnew7,li2023panoptic,liang2023efficient,addnew8,addnew9}.

Existing state-of-the-art deep learning methods for SAR ATR require a substantial amount of labeled training samples \cite{reff4,reff5,reff6, addnew10}.
However, it is a significant challenge for most SAR applications to collect and annotate a sufficient amount of SAR images \cite{reff7,reff8, isprs1}.
Existing ATR methods face degradation in performance when faced with limited SAR training data \cite{add1, add2, add3,isprs5}.
This issue has recently received considerable attention in research, known as SAR ATR with limited training data \cite{r1,r2,r3,r4,r5}. 

Generalization is currently the core contradiction and obstacle of SAR ATR with limited training data.
Some SAR ATR methods have improved the model's generalization under limited training data by employing targeted techniques, such as data augmentation and specialized modules \cite{r6,r7,r8,r9,r10}.
Despite the vigorous development of this field, we find that the fundamental theory is still missing. The critical question is still leaved unanswered: What problems does the limited SAR data cause, resulting in the weak generalization of ATR models? 

\begin{figure*}[!htb]
\begin{center}
\subfigure[ATR with Ample SAR Data]{\label{1.2}\includegraphics[width=0.33\textwidth]{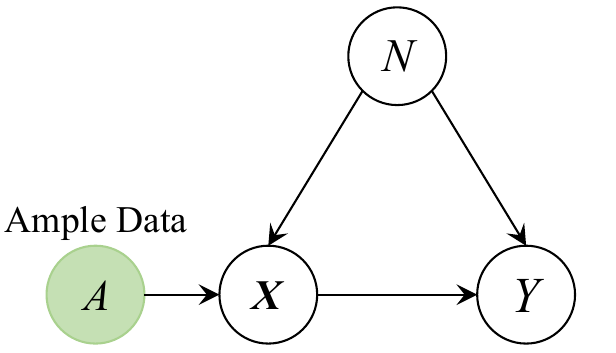}}
\subfigure[ATR with Limited SAR Data]{\label{1.3}\includegraphics[width=0.33\textwidth]{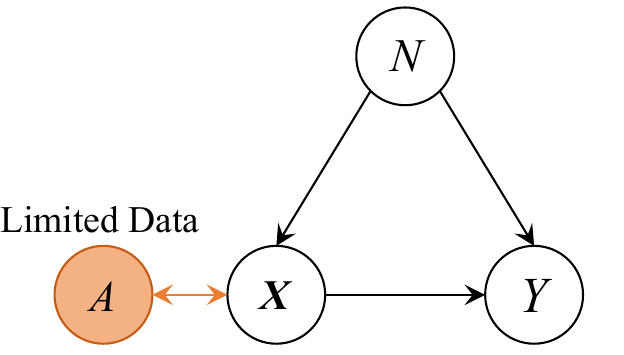}}
\subfigure[Ideal Solution with Limited SAR Data]{\label{1.4}\includegraphics[width=0.33\textwidth]{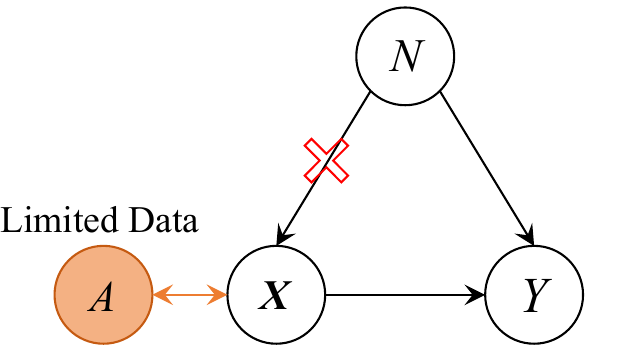}}\\
\end{center}
\caption{Structural Causal Model for SAR ATR with Ample or Limited data. The SCM indicates how related variables, target attributes $A$, features $X$, confounding noise $N$ affected with the predicted class $Y$. (a) with ample SAR data, the ATR model achieves $P(Y|{{X}}) \approx P(Y|do(x))$, (b) with limited SAR data, the ATR model faces $P(Y|{{X}}) \ncong P(Y|do(x))$, (c) an ideal solution directly modeling $P(Y|do(x))$ using backdoor adjustment.}
\label{causal}
\end{figure*}

According to the causal theory \cite{pearl2009causality, morgan2015counterfactuals, pearl2009causal}, the key to the SAR ATR generalization lies in whether the scarce SAR data can help the model achieves the true causality between the features ${X}$ and the output label $Y$.
As shown in Fig. \ref{causal} (a), when estimating the causality $P(Y|{{X}})$ between ${{X}}$ and $Y$ with ample SAR data, the causality between target attribute ${A}$ in SAR images and the feature ${{X}}$ is purely ${{A}} \to {{X}}$. 
Thus, ${{A}}$ can serve as an instrumental variable \cite{glymour2016causal} that forms a collider ${{A}} \to {{X}} \rightarrow N$ blocking all the causalities between ${{A}}$ and the noise $N$. In this way, with ample SAR data, $P(Y|{{X}}):=P(Y|{{A}}) \approx P(Y|do(x))$, meaning the model estimate the true causality only via ${{A}} \to {{X}} \to Y$. $do(\cdot)$ is the do operator in causal theory \cite{pearl2012calculus}. Therefore, with ample SAR data and true causality between $X \to Y$, the ATR model can achieve precise recognition performance and superior generalization \cite{pfister2019invariant}. 

However, with limited SAR data, the causality between ${A}$ and the feature ${{X}}$ is ${{A}} \leftrightarrow {{X}}$, not purely ${{A}} \to {{X}}$, as shown in Fig. \ref{causal}. 
The causal path ${{X}} \to {{A}}$ occurs because the model can easily establish a one-to-one mapping between scarce ${{A}}$ in the limited SAR images and their corresponding features ${{X}}$. 
When the ATR model estimates the causality between ${{X}}$ and $Y$, some spurious correlations are introduced via ${{X}} \rightarrow N \to Y$, thus, $ P(Y|{{X}}) \ncong P(Y|do(x))$.
Facing practical applications, the ATR model cannot reproduce the achieved accuracy performance on limited SAR training sets, and the recognition performance noticeably decreases with weak generalization \cite{scholkopf2012causal}.

In that case, to improve the generalization with limited SAR data, can we directly chase the true causality between ${{X}}$ and $Y$ just via ${{X}} \to Y$ by eliminating the backdoor path ${{X}} \leftarrow N \to Y$?
Fortunately, the backdoor adjustment \cite{greenland1999causal} can be employed to eliminate the effect of $N$.
In practical applications, the effectiveness of backdoor adjustment relies on an accurate estimation and effective elimination of $N$.
However, in SAR ATR, it is hard to precisely estimate and eliminate $N$ for two reasons: 

1) Inter-class variability and intrinsic-similarity confusion of $N$: $N$ is a set of multiple factors, such as clusters, shadow regions in SAR images, and so on. Not all classes are affected by the same subset of $N$ in the current constructed feature space. It is not possible to estimate $N$ for all classes using a certain fixed way. Moreover, $N$ is prone to be confused with intrinsic inter-class similarity, for example, zebras are inherently more similar to horses than to cows. The similarity generated by $N$ on the inter-class feature distribution will be confused with intrinsic inter-class similarity. 

2) Limitations of empirical risk minimization (ERM) with limited SAR data: Most of the existing optimization objects, like cross-entropy loss and contrastive learning loss, are based on ERM, and require ample SAR data to be effective. With limited SAR data, it is hard for these ERM-based loss to eliminate the effect of $N$ on ${{X}}$.
In conclusion, it is possible but hard for SAR ATR with limited data to propose a causal interventional method to precisely estimate and eliminate the effect of $N$ on ${{X}}$ for each class.

Therefore,  we further propose a causal SAR ATR with limited data via dual invariance to precisely estimates, and effectively eliminates the effect of $N$ on $X$. 
The dual invariance denotes one inner-class invariant proxy for precisely estimation of the effect of $N$, and one noise-invariance loss for effectively elimination of the effect of $N$.

1) Precisely estimation: It is necessary for the precise estimation of the effect of $N$ to require addressing the inter-class variability and intrinsic-similarity confusion of $N$. 
To accomplish this, we first propose an inner-class invariant proxy to estimate each class center. Then, a normalized similarity measure is calculated based on the invariant proxy of each class to remove the intrinsic-similarity confusion when calculating inter-class similarity.
By utilizing the normalized similarity, we can accurately estimate the effect of $N$ for each class, thereby resolving the inter-class variability and intrinsic-similarity confusion of $N$.

2) Effectively elimination: Based on the reduced effectiveness of ERM-based optimization objects with limited SAR data, it is difficult to effectively eliminate the effect of $N$ using ERM. 
Therefore, inspired by invariance risk minimization  \cite{IRM,RIRM}, we propose an noise-invariance loss that transforms the data quantity requirement of ERM into a requirement for annotating the noise environment. 
By using the previously estimation of the effect of $N$ as annotations of noise environment, the noise-invariance loss automatically divides the SAR images into multiple noise environments. 
Then the noise-invariance loss optimizes the invariance of each class's features across these environments, thereby achieving effective elimination of the effect of the noise $N$ on the features ${{X}}$.

As a result, with limited SAR data, our causal ATR achieves the true causality between ${{X}}$ and $Y$ via $P(Y|do(x))$ without the spurious correlations introduced by $N$. 
It is worth noting that the estimation of inner-class invariance features with limited SAR data can be disrupted by outlier SAR images. 
Additionally, the estimation of inner-class invariance features can also be affected by local features from interference regions in SAR images. 
To address this, we propose an instance-spatial weighting module that filters and reduces the contribution of outlier SAR images to inner-class invariance features. Moreover, we weight spatial features based on their effectiveness in recognition to estimate a precise inner-class invariance feature for each class.
The innovations of our method are summarized as below.

(1) We begin with a structural causal model (SCM) and try to fundamentally analysis and explains why the limited SAR data leads to weak generalization. The proposed SCM can also provide a principled solution to improve the generalization of ATR models with limited SAR data. 

(2) We propose a dual invariance comprising the inner-class invariant proxy and the noise-invariance loss. In the context of limited SAR data, the inner-class invariant proxy facilitates precise estimation of the effect of $N$ on $X$. The noise-invariance loss transforms the data quantity requirement of ERM into the need for annotating the noise environment, thereby effectively mitigating the impact of $N$ on ${{X}}$.

(3) Our method achieves state-of-the-art performance in the recognition of MSTAR, OpenSARship and FUSAR-Ship data sets with limited training data. The method soundness verification and ablation experiments validate the effectiveness of our methods.

The remainder of this paper is organized as follows: 
The problem formulation is presented in Section \ref{ProblemFormulation}.
The details of the proposed method are presented in Section \ref{ProposedMethod}. 
The effectiveness of the proposed method is validated through experiments in Section \ref{ExperimentsResults}. 
The conclusions are drawn in Section \ref{conclusions}. The related works is introduced in \ref{Appendix1}.

\section{Problem Formulation} \label{ProblemFormulation}
In this section, we propose a structural causal model (SCM) to systematically research ATR with limited SAR data. Then a causal explanation for weak generalization of ATR with limited SAR data is presented in detail.

\subsection{Structural Causal Model}

A SCM is constructed as shown in Fig. \ref{causal}. There are four variables: Target attributes $A$, features $X$, confounding noise $N$ and predicted class $Y$. 
The SCM is a directed acyclic graph which can indicate how variables of interest {A, X, N} affect with the class predicted by ATR model. 
As follow, the underline rationale behind SCM is detailed. 

$A \to X \to Y$. 
In general, the ATR method consists of two stages to recognize:
1) ATR method first extracts low-dimensional features $X$ from target attributes $A$ in SAR images
2) then ATR method uses these features $X$ to predict the class $Y$.
Thus, an ideal ATR method should achieve the true causality between $X$ and $Y$ only via the frontdoor path $X \to Y$. 
However, given the unique characteristics of SAR images, there is an inevitable interference from noise $N$. 
The noise $N$ infiltrates the features $X$, subsequently introducing a spurious correlation into the recognition process, following the indirect path of $N \to X \to Y$.

$N \to X \to Y$. In the recognition process of SAR images, the noise $N$ is inevitably introduced for two reasons: 
1) $N$ encapsulates the background noise inherent in SAR images, encompassing elements like cluster and shadow regions. These elements disrupt the efficiency of feature extraction $X$. 
It's clear that an ATR method predicated on background noise will struggle to achieve optimal generalization. 
This circumstance underscores the efficacy of some ATR methods that use attention mechanisms to confront the limitations encountered in SAR ATR processes with limited data \cite{fslmodel4,ATR3,r3,open2}.
2) The presence of resolution units, appearing as scattering points in SAR images, combined with the sensitivity of SAR images to imaging conditions, invariably results in inner-class variations of scattering characteristics. 
Therefore, an effective SAR ATR method dealing with limited data should inherently be equipped to manage the inner-class variations of scattering characteristics. 
This observation underscores the efficiency of ATR methods with limited SAR data, that are based on contrastive learning \cite{fslmodel2,r5,reduce1,intro6,lacking2}.

At this juncture, a discerning reader might observe that the causal graph depicted in Fig. \ref{causal} is applicable not just to limited, but also to ample SAR data scenarios.
Therefore, in the next section, we use the proposed SCM to illustrate the causal explanation for weak generalization with limited SAR data. 


\subsection{Causal Perspective on Limited SAR Data Generalization}
To understand why the generalization of the ATR method experiences a significant drop with limited SAR data, compared to abundant SAR data, we initially delve into the inherent causality in recognition with ample SAR data. 
Subsequently, we explore the additional problems that lead to poor generalization in recognition when SAR data is limited.

\textbf{Ample SAR data}: The probability of having more target attributes $A$ is higher in the case of abundant SAR data compared to limited SAR data. Considering the $ith$ SAR image which includes a subset of $A$, and the corresponding extracted features $X$, it becomes challenging for the ATR model to establish a direct one-to-one mapping between $X$ and $i$. This complexity arises due to two reasons: 
1) With an abundance of SAR data, the ATR model's attempt to establish a one-to-one mapping is akin to looking for a needle in a haystack. 
2) The sheer volume of SAR data increases the likelihood of encountering SAR images that contain similar subsets of $A$. 
Therefore, when dealing with abundant SAR data, the causal relationship between $A$ and $X$ is purely $A \to X$.

Consequently, $A$ acts as an instrumental variable in the $A \to X \rightarrow N$ pathway, thereby creating a collider. This collider structure makes $A$ and $N$ independent despite being linked via $X$ \cite{tang2020long, greenland1999causal}. Therefore, by modeling $P(Y|X):=P(Y|A)$, the ATR model ensures that $N$ no longer influences $Y$, i.e., $P(Y|X) \approx P(Y|do(X))$, and attains accurate recognition performance coupled with strong generalization, as depicted in Fig. \ref{causal}(a). 
It's worth noting that $P(Y|do(X))$ represents the ideal ATR method which effectively "cuts off" the backdoor pathway $N \to X$.

\textbf{Limited SAR data}: However, when dealing with limited SAR data, the pathway $X \to A$ occurs due to the relative ease for the ATR method to establish a one-to-one mapping between $X$ and $i$, as illustrated in Fig. \ref{causal}(b). Under these conditions, $P(Y|X) \not\approx P(Y|do(X))$ via the path $X \rightarrow N \to Y$. 
Consequently, in the context of limited SAR data, the noise $N$ induces spurious correlations in the ATR method, compromising its generalization capability.

For the true causality between $X$ and $Y$ which is just via $X \to Y$, and for the improvement of generalization, we propose a causal SAR ATR method with limited data via dual invariance. This method precisely estimates and effectively eliminates the effect of $N$ on $X$, as shown in Fig. \ref{causal}(c). The proposed solution will be presented in detail in the following sections. More details are formally showed in \ref{Appendix2}.

\begin{figure*}[htb]
\centering
\includegraphics[width=0.95\textwidth]{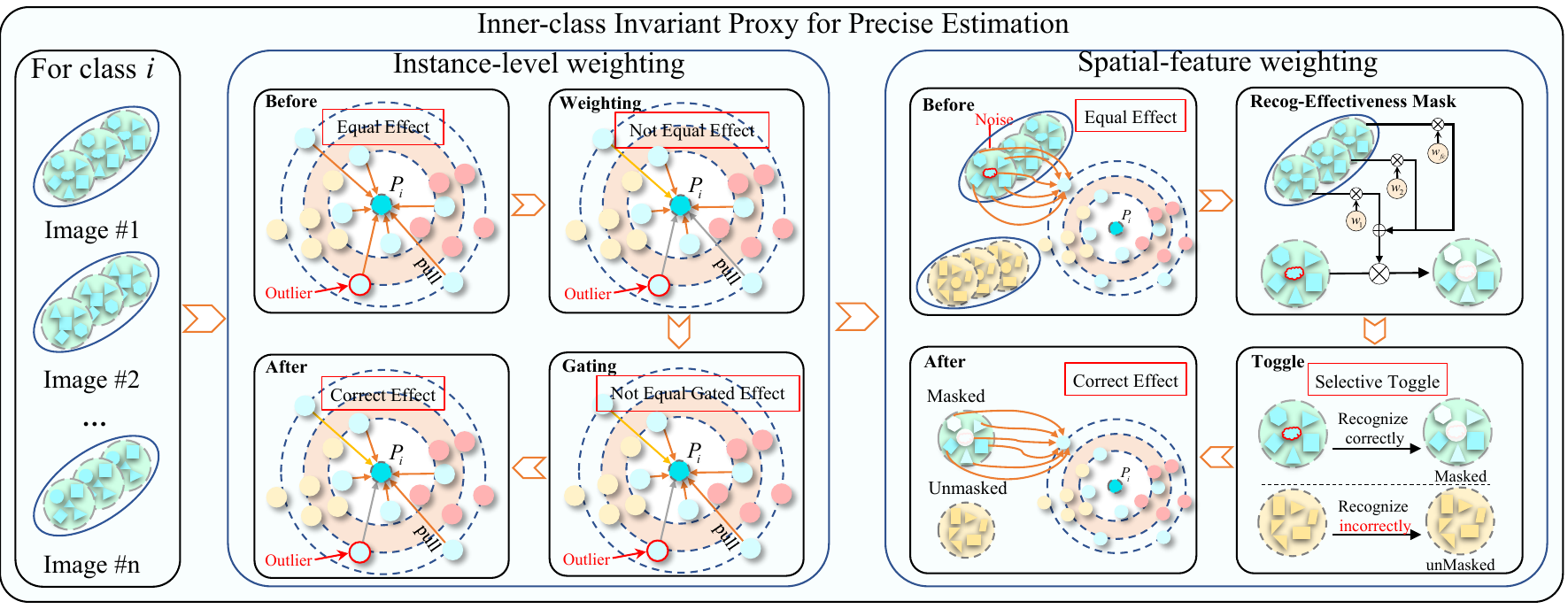}
\caption{The process of the inner-class invariant proxy for precise estimation of the effect of $N$ on $X$. To tackle with outlier SAR images and features from noise regions, an instance-spatial weighting module is proposed to filters and reduces the contribution of outlier SAR images to inner-class invariance features. Then, the instance-spatial weighting module weights spatial features based on their effectiveness in recognition to estimate a precise inner-class invariance features.}
\label{IIP}
\end{figure*}

\section{Proposed Method}\label{ProposedMethod}

Due to the introduction of spurious correlation by $N$, the causality between $X$ and $Y$ is compromised, resulting in weak generalization of ATR with limited SAR data.
In this section, we propose a causal solution using dual invariance to precisely estimate and effectively eliminate the effect of $N$ on $X$ for improving generalization of ATR with limited SAR data.
Then, we provide a detailed description of the dual invariance approach, explaining how it precisely estimates and effectively eliminates the effect of $N$.

\subsection{Causal Solution}

In this section, we first provide a confounding case with limited SAR data, then a causal intervention is introduced in detail by implementing the causal intervention $P(Y|do(X))$, as shown in Fig. \ref{causal}(c).

Based on the law of total probability, the confounding case with the spurious correlation introduced by $N$ can be presented as
\begin{equation}
P(Y|X)=\sum_N P(Y|X,N)P(N|X)
\end{equation}
which include the true causality between $X$ and $Y$ via $X \to Y$, and the spurious correlation path via $X\to N\to Y$, as shown in Fig. \ref{causal}(b).
The spurious correlation introduced by $N$ can be eliminated by using the conventional backdoor adjustment \cite{pearl2009causal, qin2021causal}.

\textbf{Definition 1}: (The Backdoor Criterion) Given a pair of variables $(X, Y)$ in a directed acyclic graph $G$, a variable $Z$ satisfies the backdoor criterion with respect to $(X, Y)$, if (i) no node in $Z$ is a descendant of $X$, and (ii) $Z$ blocks every path between $X$ and $Y$ which contains an arrow into $X$.

Therefore, in our SCM in Fig. \ref{causal}(c), $N$ satisfies The Backdoor Criterion for $(X, Y)$, backdoor adjustment is to replace $P(N|X)$ with $P(N)$, which yields the true causal effect of $X on Y$, i.e., mitigates the confounded effect of $N$. the correlation between X and Y in the conventional classifier can be formulated as
\begin{equation}
P(Y|do(X)) = \sum_N P(Y|X, N)P(N)
\end{equation}
Thus, if $N$ can be estimated precisely and eliminated, the ATR method with limited SAR data can achieve accurate recognition performance and strong generalization. A more detailed proof is listed in \ref{Appendix3}.

However, in SAR ATR, it is hard to precisely estimate and eliminate $N$ when directly using backdoor adjustment for two reasons:

1) Difficulty in precise estimation of $N$. 
In SAR images, in addition to the region containing target attribute information $A$, there are also noise areas, such as cluster and shadow regions. These regions affect the extracted features $X$, and subsequently impact the predicted class $Y$. Moreover, even regions containing target attribute information $A$ also mixed noise, such as speckle. 
Thus, $N$ contains many factors that affect the features $X$. Not all classes are affected by the same subset of $N$ in the currently constructed feature space. It is not possible to estimate $N$ for all classes using a certain fixed way, i.e., the inter-class variability of $N$ as mentioned in the introduction.

Furthermore, attribute information $A$ of different targets inherently has some similarities. Insisting on the same separability among the features of different targets may disrupt the original effective feature distribution \cite{treisman1980feature,deng2019arcface,dy2004feature}, i.e., the intrinsic-similarity confusion of $N$ as mentioned in the introduction.

2) Difficulty in effective elimination of $N$.
Most prevalent optimization objectives, such as cross-entropy loss and contrastive learning loss, are fundamentally grounded in ERM. This means that these objectives strive to minimize the average loss over a given set of training samples. They operate effectively when provided with ample SAR data, as the ERM framework fundamentally assumes that the training data is a representative sample of the entire population \cite{vapnik1991principles, chaudhuri2011differentially}.

However, in contexts where SAR data is limited, these ERM-based loss functions encounter difficulties, i.e., a less accurate representation of noise $N$. Therefore, when ERM-based loss functions attempt to optimize based on this limited SAR data, they struggle to effectively mitigate the influence of $N$ on $X$. Consequently, this can negatively affect the overall modeling of $P(Y|do(X))$.

Therefore, in the endeavor to model $P(Y|do(X))$, we propose a further effective implement, introducing a causal SAR ATR approach for limited data via dual invariance. This includes one inner-class invariant proxy for the precise estimation of the effect of $N$, and one noise-invariance loss for the effective elimination of the effect of $N$.  The specifics of the dual invariance are delineated in the following sections.

\begin{figure*}[htb]
\centering
\includegraphics[width=0.95\textwidth]{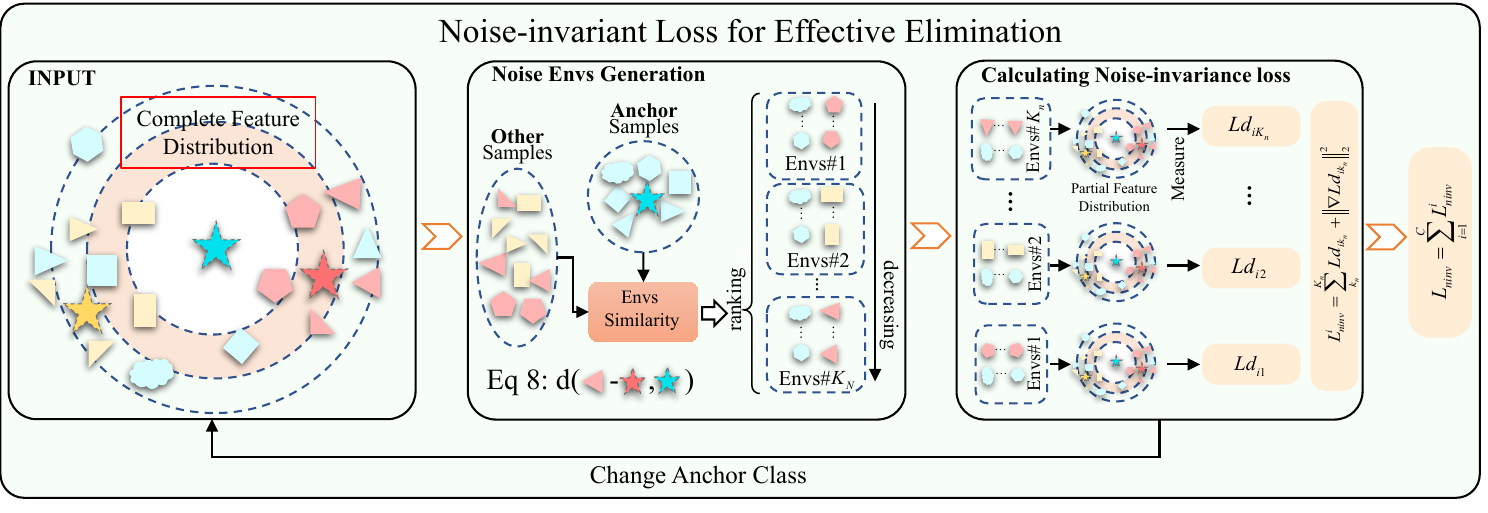}
\caption{The process of the noise-invariant loss for effective elimination of the effect of $N$ on $X$. The noise-invariance loss sets each class as an anchor class individually, and carry out virtual dependence measurement, which leverly utilizes the previous accurate estimation of $N$ as the annotation of the noise environment. Finally, the noise-invariance loss is calculated to effectively eliminate the the effect of $N$ on $X$.}
\label{NIL}
\end{figure*}

\subsection{Dual Invariance}
In this section, we delve into the specifics of the proposed dual invariance which consists of inner-class invariance proxy and noise-invariance loss. We illustrate how the inner-class invariance proxy accurately estimates the influence of $N$ on $X$ for each individual class. 
In addition, we explain how the noise-invariance loss changes the data quantity requirement intrinsic to ERM into a requirement for annotating the noise environment. This transformation allows for the effective elimination of the impact of $N$ on $X$.

\subsubsection{Inner-class Invariant Proxy}

To accurately estimate the effect of $N$ on $X$ under limited SAR data, as mentioned earlier, two problems must be resolved, namely inter-class variability and intrinsic-similarity confusion of $N$. Both problems can be solved by obtaining an accurate inner-class invariant proxy for each class, as shown in Fig. \ref{IIP}.

Suppose the inner-class invariance proxy for each class is $\{P_1, P_2,…, P_C\}$, where $C$ is the number of classes,
The solution to the inter-class variability and intrinsic-similarity confusion of $N$ is as follows:
Firstly, for the n samples $\{{\bf{x}}_{i1},..,{{\bf{x}}_{iK}}\}$ and corresponding features $\{f_{i1},…,f_{iK}\}$ of the $i$-th class, we take the inner-class invariant proxy as the anchor for the $i$-th class. 
The distance between $f_{ik}$ and the anchor, $d(f_{ik},P_i)$ acts as an estimate of the impact of $N$ on the inner-class feature distribution, thereby solving the inter-class variability of $N$.
Then, for the intrinsic-similarity confusion of $N$, regarding $f_{ik}$ and the $j$-th class, $d(f_{ik}-P_i,P_j)$ as an estimate of the impact of $N$ on the inter-class feature distribution between the $i$-th class and the $j$-th class. 
This avoids the influence of the intrinsic-similarity confusion of the effect of $N$ on $X$.

Therefore, under limited SAR data, it is necessary to accurately estimate the inner-class invariant proxy. To mitigate the impact of outlier SAR images and features from noise regions in the SAR image on the inner-class invariant proxy, we propose an instance-spatial weighting module. This module filters and reduces the contribution of outlier SAR images to inner-class invariance features. Moreover, we weight spatial features based on their effectiveness in recognition to estimate a precise inner-class invariance feature for each class. The specific pipeline can be summarized as follows.

Given the initialized inner-class invariance proxy $\{P_1, P_2,…, P_C\}$ for all classes, the features $\{f_{i1},…,f_{iK}\}$ of the $i$-th class, and the current training step $t \ge 0$.

Step 1, instance weighting. Under the training step $t$, first the distance between each feature $f_{ik}$ and $P_i$ is calculated, denoted as $d^t(f_{ik},P_i)$.
Then the impact of each sample on $P_i$ is adjusted through a dynamic parameter $\lambda_{iw}^k$. Therefore, it is organized into an optimization loss: 
\begin{equation}
L_p^i= - \sum_k^K \lambda_{iw}^k *d^t\left (l2n\left (f_{ik}\right ),l2n\left (P_i\right )\right )
\end{equation}
where $l2n(\cdot)$ is the L2 normalization, $d^t(\cdot)$ is the cosine similarity for scale invariance \cite{cos4}, and $\lambda_{iw}^k$ is calculated as 
\begin{equation}
\lambda_{iw}^k = \left (1-\beta * \frac{d^t\left (l2n\left (f_{ik}\right ),l2n\left (P_i\right )\right )+2}{2} \right )^{\rho }
\end{equation}
where $\rho \ge 0$ is a parameter to adjust the $\lambda_{iw}^k$, and $\beta = 1$ if $\frac{d^t\left (l2n\left (f_{ik}\right ),l2n\left (P_i\right )\right )-d^{t-1}\left (l2n\left (f_{ik}\right ),l2n\left (P_i\right )\right )}{d^t\left (l2n\left (f_{ik}\right ),l2n\left (P_i\right )\right )} \ge \epsilon$  else 0, serving as a gate. The above $\lambda_{iw}^k$ corrects the contribution of outlier SAR samples to the inner-class invariance proxy according to the quality of the sample.

Step 2, spatial feature weighting. For $f_{ik}$, we calculate a mask based on the recognition effectiveness of $f_{ik}$ to re-weight $f_{ik}$ spatially:
\begin{equation}
    f^w_{ik} =(1+\alpha *(M_{ik}-1)) \odot f_{ik}
\end{equation}
where $\odot$ represents the Hadamard product, $\alpha= 1$ if $\arg\max(Pred(x_{ik}))=i$ else $0$, $Pred(x_{ik})$ is the predicted class of the sample $x_{ik}, \arg\max(\cdot)$ is the corresponding index of the maximum value, and $M_{ik}$ is the mask, which can be calculated as 
\begin{equation}
M_{ik}=\sum_c^C w_{ik}(c)*f_{ik}(c)
\end{equation}
where $f_{ik}(c)$ is the feature maps in the $c$-th channel of $f_{ik}$, and $W_{ik}=\{w_{ik}(1),…, w_{ik}(C)\}$ is obtained by retrieving from $W_{fc}$ through the index $\arg\max(Pred(x_{ik}))$, i.e., $W_{ik}=W_{fc} [\arg\max(Pred(x_{ik}))]$, where $W_{fc}$ is the last dense layer in the classifier. 
By using the above method, the contribution of features from the noisy region in the SAR image can be reduced, thus obtaining an accurate estimate of the inner-class invariance proxy.

Step 3, hence, at training step $t$, a $L_p$ can be calculated to achieve an accurate inner-class invariant proxy: 

\begin{equation}
    L_p =- \sum_i^C \sum_k^K \lambda_{iw}^k *d^t(l2n(f^w_{ik}),l2n(P_i))
\end{equation}

Through the above method, we have corrected the contribution of outlier SAR samples and noisy spatial features to the proxy, obtained accurate inner-class invariant features. 
For any feature $f_{ik}$, the influence of $N$ can be estimated in inner-class and inter-class feature distribution through $d^t(f_{ik},P_i)$ and $d^t(f_{ik}-P_i,P_j)$, thus achieving the precise estimation of the effect of $N$ on $X$. 
Next, the noise-invariance loss for the effective elimination of the effect of $N$ on $X$ is presented in detail.

\subsubsection{Noise-Invariance Loss}
While we have achieved precise estimation of the effect of $N$ on $X$, most existing optimization objectives are based on ERM which requires ample SAR data, and is not effective under limited SAR data. 
Therefore, inspired by the concept of IRM, we propose a noise-invariance loss, which transforms the data quantity requirement of ERM into a requirement for annotating the noise environment. 
This cleverly utilizes the previous accurate estimation of $N$ as the annotation of the noise environment. 
We set each class as an anchor class individually, carry out virtual dependence measurement, generate dependence environment, and finally calculate the noise-invariance loss, as shown in Fig. \ref{NIL}.
The specific pipeline can be summarized as follows.

Given the optimized inner-class invariance proxy $\{P_1, P_2,…, P_C\}$ for all classes, the features $D_{\lnot i}=\{f_{11},…,f_{CK}\}$ of all SAR samples not belonging to the $i$-th class, and the feature set of the $i$-th class $\mathbf{F_i}=\{ f_{i1},…,f_{iK}\}$.

In step 1, virtual noise measurement. We set the $i$-th class as the anchor class, and for the samples not belonging to the $i$-th class, the virtual noise measurement $d_{v}(f_{jk},P_i)$ between the feature $f_{jk}$ and the invariance proxy of the anchor class is presented as follows:

\begin{equation}
    d_{v}(f_{jk},P_i)=(l2n(f_{jk})-l2n(P_j)) \odot P_i
\end{equation}

By calculating the similarity between $D_{\lnot i}$ and $P_i$ of the anchor class, we obtain a similarity list $S=\{d_{v}(f_{11},P_i),\cdots, d_{v}(f_{(i-1)1},P_i),\cdots, d_{v}(f_{(i+1)1},P_i),\cdots$ $, d_{v}(f_{CK},P_i)\}$.

Therefore, through the above formula, we can obtain the similarity between all samples not belonging to the $i$-th class and $P_i$, without breaking the intrinsic inter-class similarity. This realizes the precise estimation of the effect of $N$ on the inter-class feature distribution, thus $d_{v}(\cdot)$ can be used as the annotation of the noise environments.

In step 2, noise environment generation. Based on the virtual noise measurement $d_{v}(\cdot)$ and $S$, we can generate $K_n$ noise environments. This is done by first sorting $S$ in descending order and dividing the sorted $S$ corresponding indices into $K_n$ equal sub-lists, $S=\{s_1,\cdots,s_{K_n}\}$.

Then, the features $\{f_{11},…,f_{Cn}\}$ are also divided into corresponding $K_n$ subsets $D_{\lnot A}=\{D_{{\lnot A}1},\cdots, D_{{\lnot A}K_n}\}$ according to the indices in $S=\{s_1,\cdots,s_{K_n}\}$. These subsets $\{D_{{\lnot A}1},…, D_{{\lnot A}K_n}\}$ act as $K_n$ noise environments.

In step 3, calculating noise-invariance loss. The noise-invariance loss first calculates the relative distribution measure, $Ld_{ik}$, between all features of the $i$-th class and the features of other classes not belonging to the $i$-th class under each environment within the $K_n$ noise environments, yielding a set $\{Ld_{i1},\cdots, Ld_{i{K_n}}\}$.

\begin{equation}
    Ld_{ik_{n}}= -\sum_k^{K} \log \frac{{{ed_{v}(f_{ik},P_i)}}}{{{ed_{v}(f_{ik},P_i)}} + \sum_{f_{jk}\in D_{{\lnot A}k_{n}}} {ed_{v}(f_{jl},P_i)}}
\end{equation}
where $ed_{v}(\cdot,\cdot)= \exp{d_{v}(\cdot,\cdot)}$.
Then, for the $i$-th class, the noise-invariance loss optimizes the relative feature distribution measure of the features of the anchor class to stay invariance across the $K_n$ noise environments:
\begin{equation}
    L_{ninv}^i= \sum_{k_{n}=1}^{K_n} Ld_{ik_{n}}+|| \nabla Ld_{ik_{n}}||^2_2
\end{equation}
where $||\nabla \cdot||_2^2$ calculates the gradient penalty across the $K_n$ noise environments.
Finally, each of the $C$ classes is considered separately as an anchor class. The total noise-invariance loss is calculated as:
\begin{equation}
    L_{ninv} = \sum_{i=1}^C L_{ninv}^i
\end{equation}

By utilizing the precise estimation of the effect of $N$ on inter-class feature distribution as the annotation of noise environments, the noise-invariance loss constrains the relative feature distribution of the anchor class to remain consistent across different noise environments. Thus, the noise-invariance loss transforms the data quantity requirement of ERM into the need for annotating the interference environment, thereby effectively mitigating the effect of $N$ on $X$.

Simultaneously, we also use a cross-entropy loss as the basic recognition $L_{ce}$,
\begin{equation}
    L_{ce} = \sum_{i=1}^C \sum_{k=1}^n y_{ik} log(p(y_{ik}|x_{ik}))
\end{equation}
where $y_{ik}$ is the class label of the SAR sample $x_{ik}$. The total loss can be summarized as 
\begin{equation}
    L_{total} = L_{ce} + L_p + L_{ninv}
\end{equation}

The proposed method begin with the SCM for ATR with limited SAR data, which fundamentally analyzes that the spurious correlation introduced by the noise $N$ is the inherent problem of ATR with limited SAR data. 
Following this, our method introduced a dual invariance to precisely estimate and effectively eliminate the effect of $N$ on $X$. 
Therefore, with limited SAR data, the proposed causal ATR method obtains the true causality between $X$ and $Y$ via $P(Y|do(x))$, achieving accurate recognition performance and strong generalization.

\begin{figure}[tb]
\centering
\includegraphics[width=0.45\textwidth]{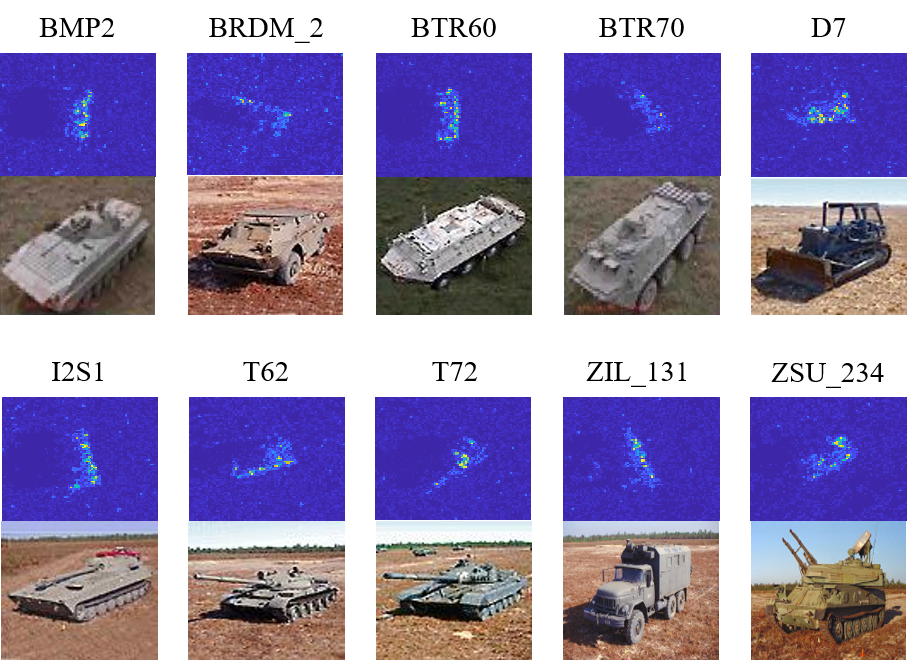}
\caption{SAR images and corresponding optical images of targets.}
\label{sampleMSTAR}
\end{figure}

\begin{figure}[tb]
\centering
\includegraphics[width=0.45\textwidth]{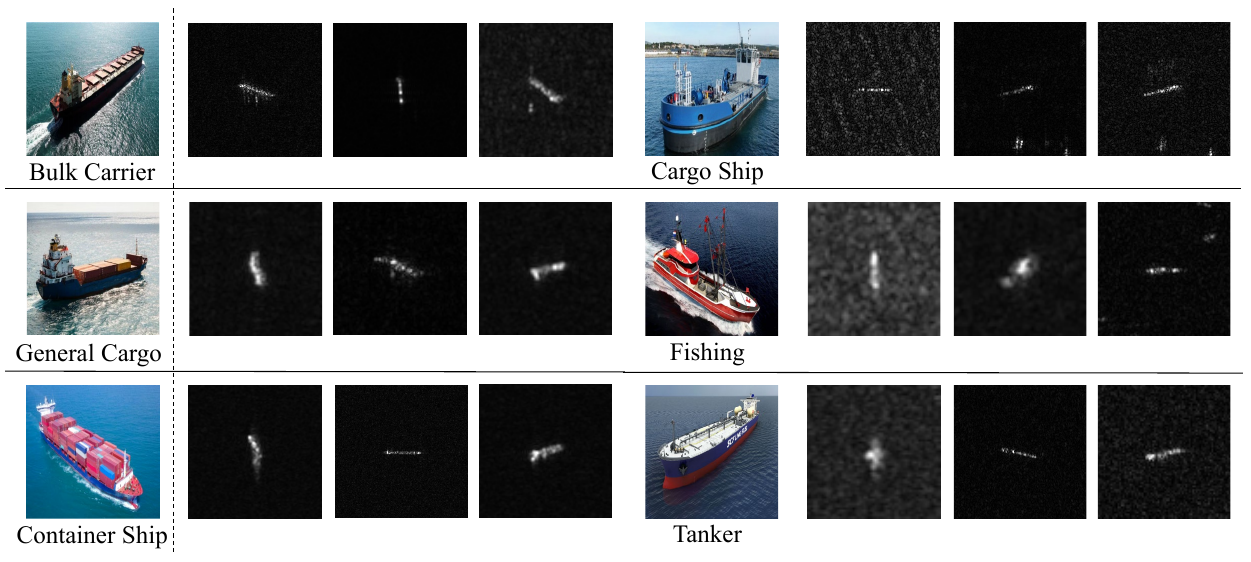}
\caption{SAR images and corresponding optical images of three-class targets in the OpenSARship dataset.}
\label{sampleOPEN}
\end{figure}

\begin{figure}[thb]
\centering
\includegraphics[width=0.99\linewidth]{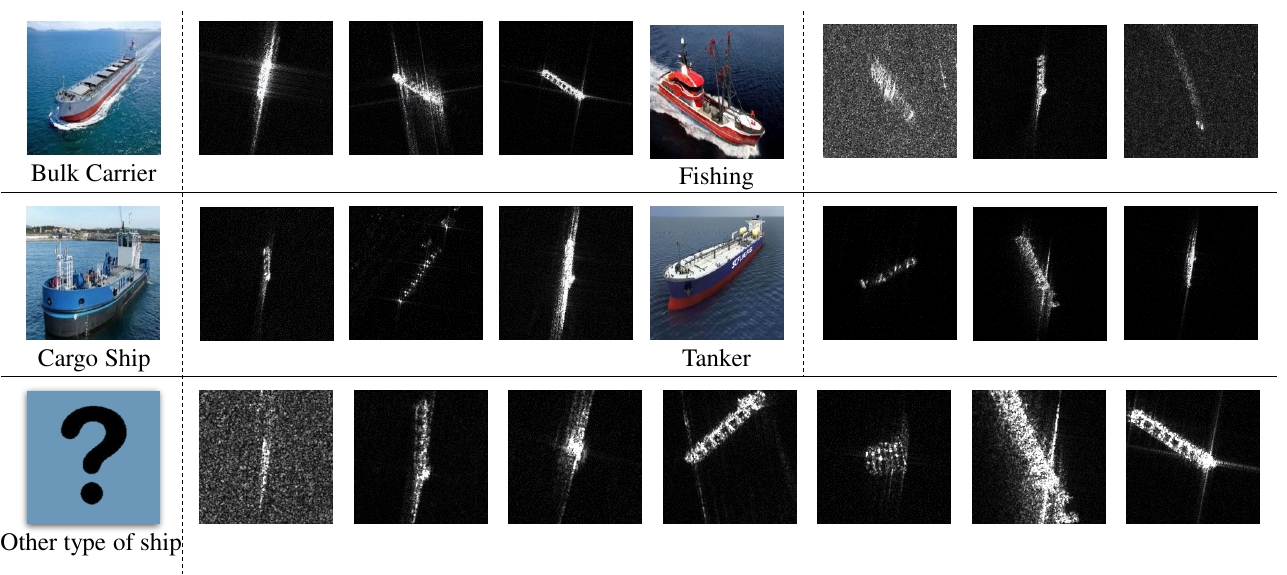} 
\caption{The FUSAR-Ship dataset includes SAR ship images and corresponding optical ship images from six ship classes. Additionally, there is a miscellaneous class called "other type of ship" that contains ships not classified into the common ship classes. This class presents a more comprehensive evaluation of the effectiveness and robustness of our method, as it involves distinguishing between similar ship types and handling overlapping features.}
\label{sampleFUSAR}
\end{figure}

\begin{table}[h]
\renewcommand{\arraystretch}{1.2}
\setlength\tabcolsep{4.4pt}
\centering
\footnotesize
\caption{Original Image Number of Different Depressions for SOC}
\label{ttnumMSTAR}
\begin{tabular}{c|cc|cc}
\toprule \toprule
\multirow{2}{*}{Class} & \multicolumn{2}{c|}{Training}            & \multicolumn{2}{c}{Testing}             \\ \cline{2-5} 
                       & \multicolumn{1}{c|}{Number} & Depression & \multicolumn{1}{c|}{Number} & Depression \\ \midrule
BMP2-9563 & \multicolumn{1}{c|}{233} & \multirow{9}{*}{$\text{17}{}^\circ$} & \multicolumn{1}{c|}{195} & \multirow{9}{*}{$\text{15}{}^\circ$} \\ \cline{1-2} \cline{4-4}
BRDM2-E71              & \multicolumn{1}{c|}{298}    &            & \multicolumn{1}{c|}{274}    &            \\ \cline{1-2} \cline{4-4}
BTR60-7532             & \multicolumn{1}{c|}{256}    &            & \multicolumn{1}{c|}{195}    &            \\ \cline{1-2} \cline{4-4}
BTR70-c71              & \multicolumn{1}{c|}{233}    &            & \multicolumn{1}{c|}{196}    &            \\ \cline{1-2} \cline{4-4}
D7-92                  & \multicolumn{1}{c|}{299}    &            & \multicolumn{1}{c|}{274}    &            \\ \cline{1-2} \cline{4-4}
2S1-b01                & \multicolumn{1}{c|}{299}    &            & \multicolumn{1}{c|}{274}    &            \\ \cline{1-2} \cline{4-4}
T62-A51                & \multicolumn{1}{c|}{299}    &            & \multicolumn{1}{c|}{273}    &            \\ \cline{1-2} \cline{4-4}
T72-132                & \multicolumn{1}{c|}{232}    &            & \multicolumn{1}{c|}{196}    &            \\ \cline{1-2} \cline{4-4}
ZIL131-E12             & \multicolumn{1}{c|}{299}    &            & \multicolumn{1}{c|}{274}    &            \\ \cline{1-2} \cline{4-4}
ZSU234-d08             & \multicolumn{1}{c|}{299}    &            & \multicolumn{1}{c|}{274}    &            \\ \bottomrule \bottomrule
\end{tabular}
\end{table}

\renewcommand{\arraystretch}{1.5}
\begin{table}[]
\centering
\scriptsize
\caption{Image Number and Imaging Conditions of Different Targets in OpenSARShip}
\label{opensarset}
\setlength\tabcolsep{2.2pt}
\begin{tabular}{c|c|ccc}
\toprule \toprule
Class          & Imaging Condition                                                                  & \begin{tabular}[c]{@{}c@{}}Training\\ Number\end{tabular} & \begin{tabular}[c]{@{}c@{}}Testing\\ Number\end{tabular} & \begin{tabular}[c]{@{}c@{}}Total\\ Number\end{tabular} \\ \midrule 
Bulk Carrier   & \multirow{6}{*}{\begin{tabular}[c]{@{}c@{}} VH and VV, C band\\  Resolution=$5-20$m\\ Incident angle=$20^{\circ}-45^{\circ}$ \\ Elevation sweep angle=$\pm 11^{\circ}$\\ ${\text{Rg20}}m \times {\text{az}}22m$\end{tabular}} & 200                                                       & 475                                                      & 675                                                    \\ \cline{1-1} \cline{3-5}
Container Ship &                                                                                    & 200                                                       & 811                                                      & 1011                                                   \\ \cline{1-1} \cline{3-5}
Tanker         &                                                                                    & 200                                                       & 354                                                      & 554                                                    \\ \cline{1-1} \cline{3-5}
Cargo          &                                                                                    & 200                                                       & 557                                                      & 757                                                    \\ \cline{1-1} \cline{3-5}
Fishing        &                                                                                    & 200                                                       & 121                                                      & 321                                                    \\ \cline{1-1} \cline{3-5}
General Cargo  &                                                                                    & 200                                                       & 165                                                      & 365     \\ \bottomrule \bottomrule                                               
\end{tabular}
\end{table}

\renewcommand{\arraystretch}{1.5}
\begin{table}[]
\centering
\scriptsize
\caption{Image Number and Imaging Conditions of Different Targets in FUSAR-Ship}
\label{fusarset}
\setlength\tabcolsep{1.5pt}
\begin{tabular}{c|c|ccc}
\toprule \toprule 
Class          & Imaging Condition                                                                  & \begin{tabular}[c]{@{}c@{}}Training\\ Number\end{tabular} & \begin{tabular}[c]{@{}c@{}}Testing\\ Number\end{tabular} & \begin{tabular}[c]{@{}c@{}}Total\\ Number\end{tabular} \\ \midrule  
Bulk Carrier          & \multirow{5}{*}{\begin{tabular}[c]{@{}c@{}}VH and VV, C band\\  Resolution=$0.5-500$m\\ Incident angle=$10^{\circ}-60^{\circ}$ \\ Elevation sweep angle=$\pm 20^{\circ}$\\ ${\text{Rg20}}m \times {\text{az}}22m$\end{tabular}} & 100                  & 173                  & 273                  \\ \cline{1-1} \cline{3-5}
Cargo Ship            &                                                                                    & 100                  & 1593                 & 1693                 \\ \cline{1-1} \cline{3-5}
Fishing              &                                                                                    & 100                  & 685                  & 785                  \\ \cline{1-1} \cline{3-5}
Other type of ship      &                                                                                    & 100                  & 1507                 & 1607                 \\ \cline{1-1} \cline{3-5}
Tanker               &                                                                                    & 100                  & 48                   & 148                  \\ \bottomrule \bottomrule                                               
\end{tabular} 
\end{table}

\section{Experiments and Results}\label{ExperimentsResults}

This section evaluates the effectiveness and robustness of our proposed method under the constraints of limited SAR training samples. We begin by introducing the MSTAR, OpenSARship, and FUSAR-Ship datasets, along with their corresponding preprocessing procedures. We then conduct method soundness experiments, consisting of ablation experiments, and visualizations of feature distribution, to validate the effectiveness of our method. Next, we present the recognition performances of our method on different datasets with limited data, showcasing its effectiveness in dealing with varying sample numbers. Lastly, we compare our proposed method to other state-of-the-art methods for ATR with limited SAR data.

\subsection{Datasets and Configuration}
To evaluate the recognition performance of our proposed method with limited SAR training samples, we utilize three benchmark datasets: the moving and stationary target acquisition and recognition (MSTAR) dataset, the OpenSARship dataset, and the FUSAR-Ship dataset. These datasets serve as valuable resources for assessing the effectiveness of our method in SAR ATR.

The MSTAR dataset, released by the Defense Advanced Research Project Agency and the Air Force Research Laboratory, is a benchmark dataset for evaluating SAR ATR performance. The data was collected using the Sandia National Laboratory STARLOS sensor platform. It consists of SAR images with 1-ft resolution in the X-band, covering a range from 0° to 360°. The dataset includes ten different types of ground targets, such as tanks, rocket launchers, armored personnel carriers, air defense units, and bulldozers. These targets exhibit variations in aspect angles, depression angles, and serial numbers. Fig. \ref{sampleMSTAR} shows the SAR and corresponding optical images of these ten target types.

The OpenSARship dataset is designed to develop advanced ship detection and classification algorithms in the presence of high interference \cite{OpenSARShip}. The dataset consists of 41 Sentinel-1 images captured under various environmental conditions. It contains a total of 11,346 ship chips, corresponding to 17 different types of SAR ships. The ship labels in this dataset are reliable as they are based on automatic identification system (AIS) information.
In our experiments, we use the ground range detected (GRD) data from Sentinel-1 IW mode, which has a resolution of $2.0m \times 1.5m$. The ships in the dataset have lengths ranging from $92m$ to $399m$ and widths ranging from 6$m $to 65$m. $Both VV and VH data are utilized in the training, validation, and testing phases of our experiments. Fig. \ref{sampleOPEN} showcases sample SAR images of a three-class target from the OpenSARship dataset.

FUSAR-Ship is another open benchmark dataset specifically designed for ship and marine target detection and recognition \cite{FUSAR}. It was compiled by the Key Lab of Information Science of Electromagnetic Waves (MoE) at Fudan University for the Gaofen-3 satellite. The Gaofen-3 satellite is the first civilian C-band fully polarized satellite-based SAR system in China, primarily used for marine remote sensing.
The FUSAR-Ship dataset serves as an open SAR-AIS matching dataset and consists of over 100 Gaofen-3 scenes, encompassing more than 5000 ship image slices with corresponding AIS information. This dataset offers distinct imaging parameters compared to the OpenSARship dataset, including incident angle, bandwidth, and resolution. The ranges of these imaging parameters in the FUSAR-Ship dataset are larger than those in the OpenSARship dataset. Although the FUSAR-Ship dataset provides a higher resolution of $0.5m$, the expanded ranges of the imaging parameters make it more challenging for recognition compared to the OpenSARship dataset.

\renewcommand{\arraystretch}{1.6}
\begin{table*}[h]
\centering
\footnotesize
\caption{Ablation Experiments: Recognition Performance (\%) of Different Ablation Configurations under Different Training Samples in MSTAR. IIP and NIL stand for inner-class invariant proxy and noise-invariance loss respectively. }
\label{ablation_tab}
\begin{tabular}{c|c|cc|ccccccccccc}
\toprule \toprule
\makebox[19pt][c]{\multirow{2}{*}{\begin{tabular}[c]{@{}c@{}}Training\\ Number\end{tabular}}} &
  \makebox[14pt][c]{\multirow{2}{*}{Method}} &
  \makebox[14pt][c]{\multirow{2}{*}{IIP}} &
  \makebox[14pt][c]{\multirow{2}{*}{NIL}} &
  \makebox[11.5pt][c]{\multirow{2}{*}{BMP2}} &
  \makebox[11.5pt][c]{\multirow{2}{*}{BRDM2}} &
  \makebox[11.5pt][c]{\multirow{2}{*}{BTR60}} &
  \makebox[11.5pt][c]{\multirow{2}{*}{BTR70}} &
  \makebox[11.5pt][c]{\multirow{2}{*}{D7}} &
  \makebox[11.5pt][c]{\multirow{2}{*}{2S1}} &
  \makebox[11.5pt][c]{\multirow{2}{*}{T62} }&
  \makebox[11.5pt][c]{\multirow{2}{*}{T72}} &
  \makebox[11.5pt][c]{\multirow{2}{*}{ZIL131}} &
  \makebox[11.5pt][c]{\multirow{2}{*}{ZSU234}} &
  \makebox[11.5pt][c]{\multirow{2}{*}{Average}} \\
                     &      &     &   &       &       &       &       &       &       &       &       &       &       &       \\ \midrule 
\multirow{4}{*}{5}  & V1   & × & × & 0.00   & 43.80 & 94.36 & 71.43 & 19.71  & 75.91 & 43.96  & 39.29 & 84.31  & 96.72  & 57.69 \\
                    & V2   & × & \checkmark  & 33.33  & 91.24 & 78.46 & 37.24 & 92.70  & 57.66 & 75.09  & 62.24 & 94.16  & 90.51  & 73.65 \\
                    & V3   & \checkmark & × & 28.72  & 83.94 & 71.28 & 43.88 & 92.70  & 67.52 & 91.94  & 61.73 & 88.32  & 95.99  & 75.34 \\
                    & Ours & \checkmark & \checkmark & 38.46  & 84.31 & 69.74 & 42.35 & 96.72  & 89.42 & 90.48  & 57.65 & 93.80  & 98.54  & 79.26 \\ \midrule 
\multirow{4}{*}{10} & V1   & × & × & 61.54  & 43.80 & 94.36 & 71.43 & 63.50  & 75.91 & 43.96  & 39.29 & 84.31  & 96.72  & 67.59 \\
                    & V2   & × & \checkmark & 69.23  & 94.16 & 54.87 & 66.84 & 93.07  & 72.26 & 91.21  & 86.73 & 92.70  & 99.27  & 83.67 \\
                    & V3   & \checkmark & × & 70.26  & 88.32 & 57.44 & 64.29 & 94.89  & 71.53 & 93.41  & 88.78 & 91.61  & 99.64  & 83.55 \\
                    & Ours & \checkmark & \checkmark & 66.15  & 95.26 & 75.38 & 80.10 & 98.91  & 94.89 & 92.67  & 96.94 & 98.18  & 100.00 & 91.18 \\ \midrule 
\multirow{4}{*}{20} & V1   & × & × & 100.00 & 83.58 & 94.36 & 71.43 & 63.50  & 75.91 & 88.28  & 39.29 & 84.31  & 96.72  & 80.16 \\
                    & V2   & × & \checkmark & 79.49  & 94.53 & 94.36 & 84.18 & 98.91  & 97.08 & 97.80  & 93.88 & 99.64  & 100.00 & 94.76 \\
                    & V3   & \checkmark & × & 26.15  & 97.81 & 87.69 & 80.61 & 88.32  & 78.47 & 92.67  & 71.43 & 98.91  & 100.00 & 84.25 \\
                    & Ours & \checkmark & \checkmark & 84.62  & 99.64 & 89.23 & 90.82 & 100.00 & 97.81 & 100.00 & 94.90 & 100.00 & 100.00 & 96.45 \\ \bottomrule \bottomrule 
\end{tabular} 
\end{table*}

The configurations of the training process and the network are presented here. The size of input SAR images is $224 \times 224$ by applying bi-linear interpolation to the original data. The values of ${K_n}$ is set as 3. 
The value of ${{\rm{margin}}}$ is set as 0.3. 
The batch size is set as 32. The learning rate is initialized as 0.01 and reduced with the 0.1 ratios for every 25 epochs. 
There are also 10 epochs to warm up for training. Other hyper-parameters are shown in Fig. \ref{IIP} and Fig. \ref{NIL}. 
The proposed method is tested and evaluated on a GPU cluster with Intel(R) Xeon(R) CPU E5-2698 v4 @ 2.20GHz, eight Tesla V100 with eight 32GB memories. The proposed method is implemented using the open-source PyTorch framework with only one Tesla V100.

\subsection{Method Soundness Verification under Constrain of Limited SAR Training Samples}

In this section, we conduct soundness verification experiments to evaluate the effectiveness of our method under the constraints of limited SAR training samples. We employ two types of experiments to validate our method:
1) Ablation experiments are performed under different configurations to assess the impact of the proposed innovations on the generalization.
2) The visualization of feature distributions is conducted to analyze the characteristics of the learned features under various configurations of the ablation experiments.

\renewcommand{\arraystretch}{1.6}
\begin{table}[h]
\centering
\footnotesize
\caption{Ablation Experiments: Recognition Performance (\%) of Different Ablation Configurations under Different Training Samples in OpenSARShip. IIP and NIL stand for inner-class invariant proxy and noise-invariance loss respectively. }
\label{ablation_tab2}
\resizebox{\linewidth}{!}{
\begin{tabular}{c|c|cc|cccc}
\toprule \toprule
\makebox[18pt][c]{\multirow{2}{*}{\begin{tabular}[c]{@{}c@{}}Training\\ Number\end{tabular}}} &
  \makebox[16pt][c]{\multirow{2}{*}{Method}} &
  \makebox[14pt][c]{\multirow{2}{*}{IIP}} &
  \makebox[14pt][c]{\multirow{2}{*}{NIL}} &
  \makebox[10pt][c]{\multirow{2}{*}{\begin{tabular}[c]{@{}c@{}}Bulk\\ Carrier\end{tabular}}} &
  \makebox[10pt][c]{\multirow{2}{*}{\begin{tabular}[c]{@{}c@{}}Container\\Ship\end{tabular}}} &
  \makebox[10pt][c]{\multirow{2}{*}{Tanker}} &
  \makebox[10pt][c]{\multirow{2}{*}{Average}} \\
                    &      &   &   &        &       &       &       \\ \midrule 
\multirow{4}{*}{10} & V1   & × & × & 4.63   & 92.23 & 76.55 & 63.48 \\
                    & V2   & × & \checkmark  & 75.79  & 55.61 & 67.80 & 64.09 \\
                    & V3   & \checkmark  & × & 9.05   & 89.77 & 66.38 & 61.34 \\
                    & Ours & \checkmark  & \checkmark  & 50.11  & 84.22 & 69.21 & 71.10 \\ \midrule 
\multirow{4}{*}{20} & V1   & × & × & 30.74  & 85.45 & 79.66 & 68.35 \\
                    & V2   & × & \checkmark  & 49.47  & 84.59 & 67.80 & 70.79 \\
                    & V3   & \checkmark  & × & 37.26  & 89.52 & 62.71 & 68.60 \\
                    & Ours & \checkmark  & \checkmark  & 53.89  & 83.23 & 82.49 & 74.57 \\ \midrule 
\multirow{4}{*}{30} & V1   & × & × & 42.32  & 78.91 & 83.62 & 69.33 \\
                    & V2   & × & \checkmark  & 100.00 & 59.19 & 67.80 & 72.87 \\
                    & V3   & \checkmark  & × & 36.21  & 87.42 & 87.85 & 72.68 \\
                    & Ours & \checkmark  & \checkmark  & 53.89  & 85.82 & 79.94 & 75.30 \\ \bottomrule \bottomrule 
\end{tabular} }
\end{table}

\begin{figure*}[!htb]
\begin{center}
\subfigure[]{\label{a1.1}\includegraphics[width=1.5in]{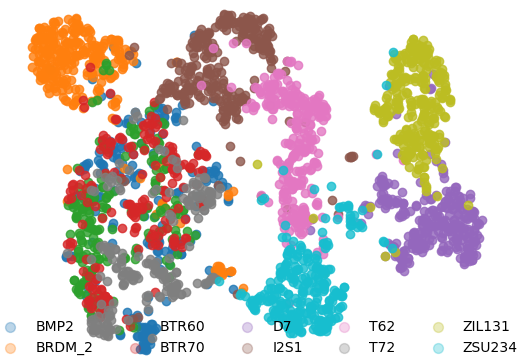}}
\subfigure[]{\label{a1.2}\includegraphics[width=1.5in]{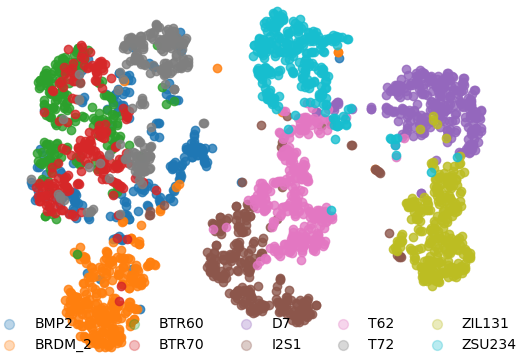}}
\subfigure[]{\label{a1.3}\includegraphics[width=1.5in]{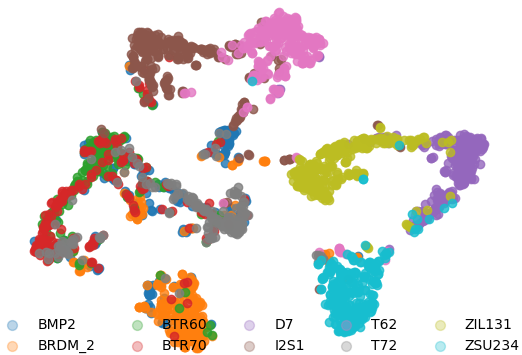}}
\subfigure[]{\label{a1.4}\includegraphics[width=1.5in]{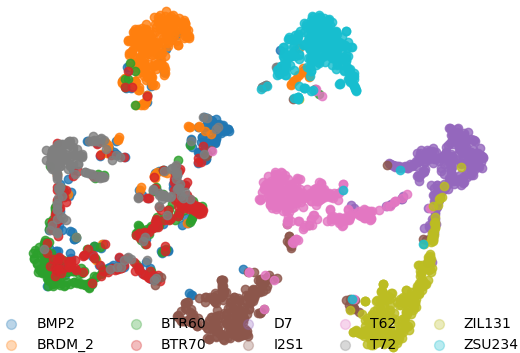}}\\
\subfigure[]{\label{a1.5}\includegraphics[width=1.5in]{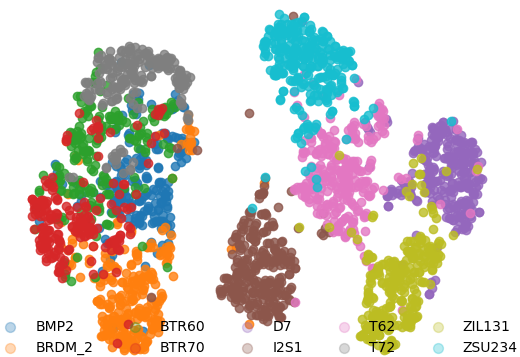}}
\subfigure[]{\label{a1.6}\includegraphics[width=1.5in]{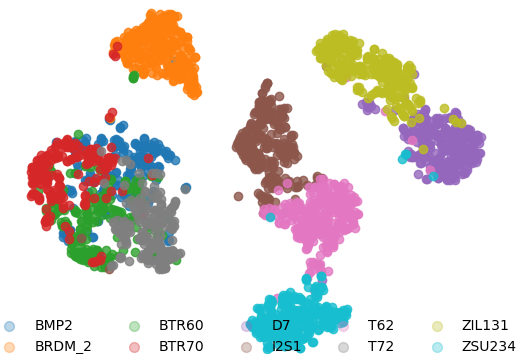}}
\subfigure[]{\label{a1.7}\includegraphics[width=1.5in]{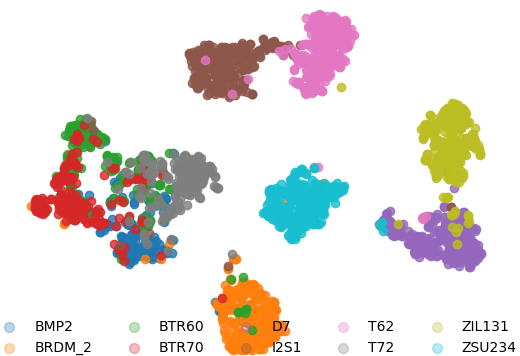}}
\subfigure[]{\label{a1.8}\includegraphics[width=1.5in]{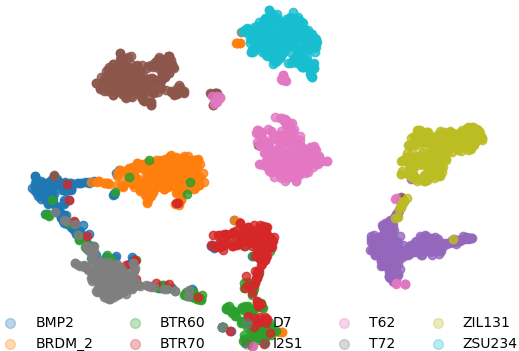}}\\
\subfigure[]{\label{a1.9}\includegraphics[width=1.5in]{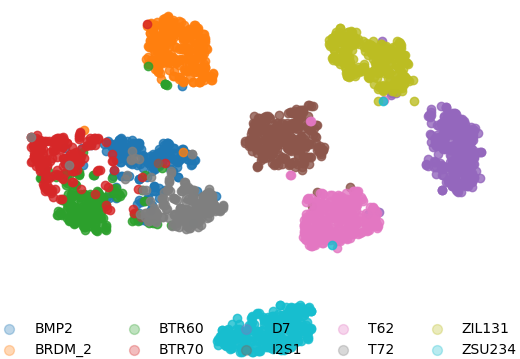}}
\subfigure[]{\label{a1.10}\includegraphics[width=1.5in]{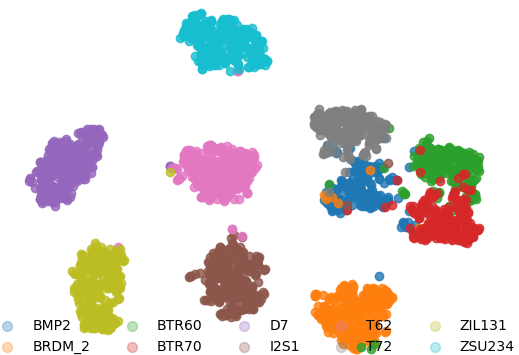}}
\subfigure[]{\label{a1.11}\includegraphics[width=1.5in]{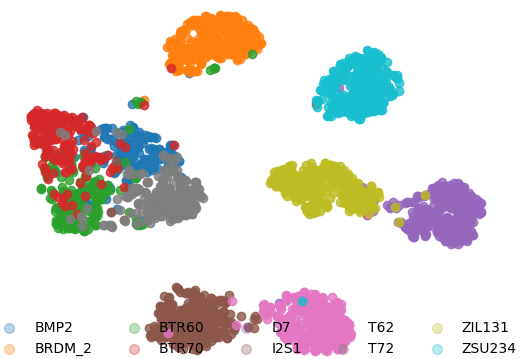}}
\subfigure[]{\label{a1.12}\includegraphics[width=1.5in]{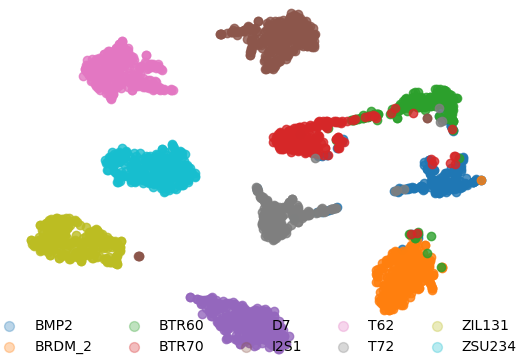}}\\
\end{center}
\caption{Ablation Experiments: Feature Visualization under Different Ablation Configurations in MSTAR. (a), (b), (c), and (d) are feature visualizations of V1, V2, V3, and the full version of our model under 5 training samples in each class, respectively. (e), (f), (g), and (h) are under 10 training samples in each class; (i), (j), (k), and (l) are under 20 training samples in each class.}
\label{ablation_fig1}
\end{figure*}

\begin{figure}[!htb]
\begin{center}
\subfigure[]{\label{a2.1}\includegraphics[width=0.22\textwidth]{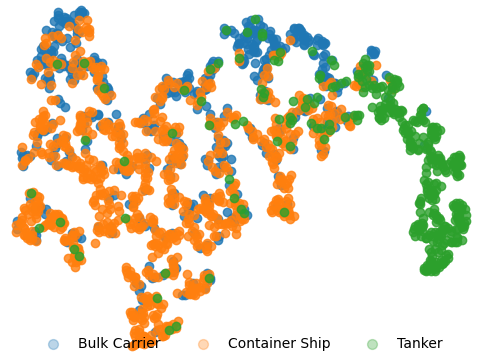}}
\subfigure[]{\label{a2.2}\includegraphics[width=0.22\textwidth]{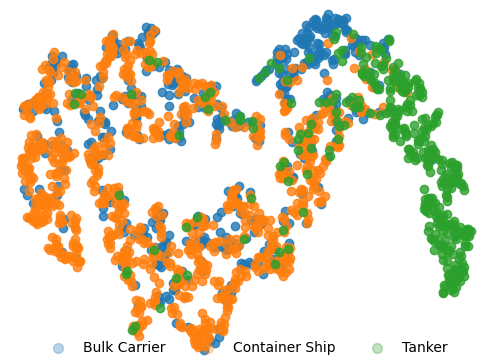}}
\subfigure[]{\label{a2.3}\includegraphics[width=0.22\textwidth]{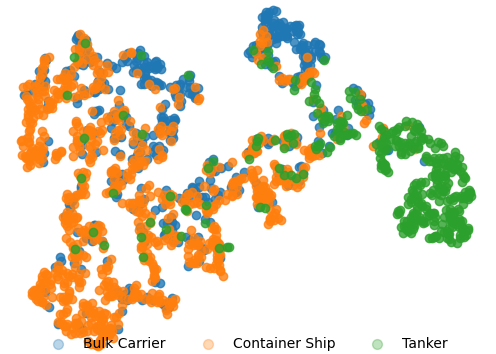}}
\subfigure[]{\label{a3.4}\includegraphics[width=0.22\textwidth]{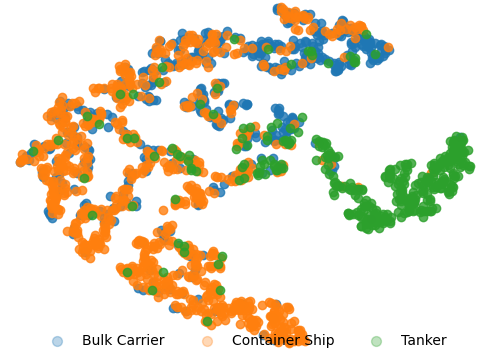}}\\
\end{center}
\caption{Ablation Experiments: Feature Visualization under Different Ablation Configurations in OpenSARShip. (a), (b), (c), and (d) are feature visualizations of V1, V2, V3, and the full version of our model under 30 training samples in each class, respectively.}
\label{ablation_fig2}
\end{figure}

\subsubsection{Ablation Experiments of Different Configurations}

We conducted ablation experiments to evaluate different configurations of our method under the limitations of limited SAR training samples under MSTAR and OpenSARShip datasets. Table \ref{ablation_tab} and Table \ref{ablation_tab2} summary the four configurations: V1, V2, V3, and Ours. 
V1 represents the vanilla model without the dual invariance. 
V2 represents the vanilla model with the noise-invariance loss and a prototype rather than the inner-class invariance proxy.
V3 represented the model with the inner-class invariance proxy and a contrastive loss rather than the noise-invariance loss.
Lastly, "Ours" refers to the full version of our causal model. 
The recognition performances of these configurations can be seen in Table \ref{ablation_tab} and Table \ref{ablation_tab2}.

From the comparison of the performance of four configurations under MSTAR and OpenSARShip dataset, 1) For MSTAR, compared V2 with V1 and V3, it is clear that when the training samples are critically limited, like 5 or 10 each class, the inner-class invariance proxy can better help the model reduce the effect of $N$ on $X$ than the noise-invariance loss. But when the training samples each class are 20, the noise-invariance loss works better than the inner-class invariance proxy. For OpenSARShip, the . This phenomenon may be caused by the complexity of different datasets, the complexity of MSTAR dataset is a litter easy than the OpenSARShip dataset, thus, the inner-class invariant proxy works better.
2) For MSTAR and OpenSARShip datasets, compared Ours with V2 and V3, the cooperation of the dual invariance can obviously improves the generalization of the model. 
From the ablation experiments, it has illustrated that not only can individual invariance paired with simple methods effectively obtain the true causality between $X$ and $Y$, but also the cooperation of dual invariance can achieve better generalization.

Therefore, through the ablation experiments with four configurations of our methods under the constrains of the limited SAR training samples, it has been validated that our method and innovations have an effective improvement of the model's generalization.

\renewcommand{\arraystretch}{1.5}
\begin{table*}[h]
\centering
\footnotesize
\caption{Recognition Performances of 10 Classes under Different Training Data in MSTAR}
\label{mstar-results}
\setlength{\tabcolsep}{1.5mm}{
\begin{tabular}{c|ccccccccccccc}
\toprule \toprule 
\multirow{2}{*}{Class} & \multicolumn{10}{c}{Training Number in Each Class}                                                                              \\ \cline{2-11} 
                       & 2       & 5      & 10      & 20      & 25      & 30      & 40      & 60      & 80      & 100     \\ \midrule 
BMP2                                       & 18.46\% & 38.46\% & 66.15\%  & 84.62\%  & 85.64\%  & 97.95\%  & 97.44\%  & 97.95\%  & 98.97\%  & 100.00\% \\
BRDM2                                      & 68.98\% & 84.31\% & 95.26\%  & 99.64\%  & 98.18\%  & 98.18\%  & 98.91\%  & 100.00\% & 100.00\% & 100.00\% \\
BTR60                                      & 66.15\% & 69.74\% & 75.38\%  & 89.23\%  & 94.36\%  & 96.41\%  & 92.31\%  & 96.92\%  & 97.44\%  & 96.92\%  \\
BTR70                                      & 20.92\% & 42.35\% & 80.10\%  & 90.82\%  & 97.45\%  & 92.86\%  & 98.47\%  & 98.98\%  & 98.98\%  & 98.98\%  \\
D7                                         & 85.04\% & 96.72\% & 98.91\%  & 100.00\% & 98.91\%  & 100.00\% & 99.64\%  & 100.00\% & 100.00\% & 100.00\% \\
2S1                                        & 62.41\% & 89.42\% & 94.89\%  & 97.81\%  & 96.35\%  & 97.45\%  & 98.54\%  & 98.91\%  & 99.27\%  & 100.00\% \\
T62                                        & 69.60\% & 90.48\% & 92.67\%  & 100.00\% & 99.27\%  & 99.63\%  & 100.00\% & 100.00\% & 100.00\% & 100.00\% \\
T72                                        & 50.00\% & 57.65\% & 96.94\%  & 94.90\%  & 96.94\%  & 97.96\%  & 98.98\%  & 100.00\% & 100.00\% & 100.00\% \\
ZIL131                                     & 72.63\% & 93.80\% & 98.18\%  & 100.00\% & 100.00\% & 99.27\%  & 100.00\% & 100.00\% & 100.00\% & 100.00\% \\
ZSU235                                     & 52.19\% & 98.54\% & 100.00\% & 100.00\% & 100.00\% & 99.27\%  & 100.00\% & 99.64\%  & 100.00\% & 100.00\% \\ \midrule 
Average                                    & 58.93\% & 79.26\% & 91.18\%  & 96.45\%  & 97.11\%  & 98.10\%  & 98.64\%  & 99.34\%  & 99.55\%  & 99.67\%  \\ \bottomrule \bottomrule
\end{tabular} }
\end{table*}

\renewcommand{\arraystretch}{1.5}
\begin{table*}[h]
\centering
\footnotesize
\caption{Recognition Performances of 3 Classes under Different Training Data in OpenSARShip}
\label{opensarship3-r}
\begin{tabular}{c|cccccccccc}
\toprule \toprule 
\multirow{2}{*}{Class} & \multicolumn{10}{c}{Training Number in Each Class}                                       \\ \cline{2-11} 
                       & 10      & 20      & 30      & 40    &50  & 60      & 70      & 80      & 100     & 200     \\ \midrule 
Bulk Carrier   & 50.11\% & 53.89\% & 53.89\% & 67.16\% & 61.05\% & 69.68\% & 70.32\% & 67.37\% & 75.79\% & 75.37\% \\
Container Ship & 84.22\% & 83.23\% & 85.82\% & 77.31\% & 87.92\% & 83.11\% & 82.49\% & 87.92\% & 86.56\% & 91.25\% \\
Tanker                 & 69.21\% & 82.49\% & 79.94\% & 83.62\% & 81.92\% & 83.90\% & 85.31\% & 81.64\% & 83.62\% & 85.03\% \\ \midrule 
Average                & 71.10\% & 74.57\% & 75.30\% & 75.73\% & 78.84\% & 79.39\% & 79.57\% & 80.61\% & 82.80\% & 85.30\% \\ \bottomrule \bottomrule 
\end{tabular} 
\end{table*}

\renewcommand{\arraystretch}{1.5}
\begin{table*}[h]
\centering
\footnotesize
\caption{Recognition Performances of 6 Classes under Different Training Data in OpenSARShip}
\label{opensarship6-r}
\begin{tabular}{c|ccccccccc}
\toprule \toprule 
\multirow{2}{*}{Class} & \multicolumn{9}{c}{Training Number in Each Class}                                       \\ \cline{2-10} 
                       & 10      & 20      & 30      & 40      & 60      & 70      & 80      & 100     & 200     \\ \midrule 
Bulk Carrier                               & 55.37\% & 57.26\% & 41.47\% & 46.74\% & 66.74\% & 52.42\% & 54.32\% & 60.63\% & 60.63\% \\
Container Ship                             & 66.95\% & 72.01\% & 81.50\% & 83.60\% & 81.50\% & 83.85\% & 80.39\% & 89.89\% & 85.82\% \\
Tanker                                     & 39.55\% & 50.28\% & 53.39\% & 39.55\% & 51.13\% & 60.73\% & 56.21\% & 52.54\% & 56.78\% \\
Cargo                                      & 35.01\% & 44.17\% & 42.37\% & 53.68\% & 40.39\% & 49.19\% & 55.66\% & 45.96\% & 60.14\% \\
Fishing                                    & 91.74\% & 84.30\% & 87.60\% & 90.08\% & 90.91\% & 88.43\% & 83.47\% & 91.74\% & 92.56\% \\
General Cargo                              & 15.15\% & 30.91\% & 51.52\% & 26.67\% & 48.48\% & 36.97\% & 41.82\% & 36.97\% & 44.24\% \\ \midrule 
Average                                    & 51.43\% & 57.71\% & 59.36\% & 60.09\% & 63.39\% & 63.87\% & 64.00\% & 65.69\% & 68.67\%  \\ \bottomrule \bottomrule 
\end{tabular} 
\end{table*}

\renewcommand{\arraystretch}{1.5}
\begin{table}[]
\centering
\footnotesize
\caption{Recognition Performances of 5 Classes under Different Training Data in FuSAR-Ship}
\label{fusar-r}
\setlength{\tabcolsep}{0.4mm}{
\begin{tabular}{c|cccccc}
\toprule \toprule 
\multirow{2}{*}{Class} & \multicolumn{6}{c}{Labeled Number in Each   Class}        \\ \cline{2-7} 
                       & 20      & 30      & 40      & 60      & 70      & 100     \\ \midrule
BulkCarrier                                & 41.47\% & 35.96\% & 49.87\% & 48.45\% & 53.07\% & 47.02\% \\
CargoShip                                  & 67.61\% & 69.80\% & 70.58\% & 74.66\% & 70.08\% & 77.63\% \\
Fishing                                    & 43.04\% & 47.27\% & 45.51\% & 47.73\% & 46.37\% & 48.42\% \\
Other type of ship                         & 96.97\% & 97.58\% & 96.93\% & 98.37\% & 96.38\% & 96.89\% \\
Tanker                                     & 23.24\% & 26.78\% & 24.32\% & 15.85\% & 26.13\% & 12.01\% \\ \midrule 
Average                                    & 64.50\% & 65.06\% & 67.32\% & 68.68\% & 68.88\% & 70.07\% \\ \bottomrule \bottomrule 
\end{tabular} }
\end{table}

\subsubsection{Feature visualizations of different configurations}

Furthermore, we visually compare the feature distributions under the four different configurations to gain a more intuitive understanding of our approach within the limitations of limited SAR training samples. The four configurations depicted in Fig. \ref{ablation_fig1} and Fig. \ref{ablation_fig2} correspond to those described in Table \ref{ablation_tab} and Table \ref{ablation_tab2}. 
After training the models, we apply them to the testing data to obtain the final features for each class. We then utilize t-SNE to visualize the feature distributions, as illustrated in Fig. \ref{ablation_fig1} and Fig. \ref{ablation_fig2}.

By comparing four configurations in Fig. \ref{ablation_fig1}, the inner-class variant proxy can help the model obtain higher inner-class compactness than the noise-invariance loss. It explains the reasons that the inner-class variant proxy works better than the noise-invariance loss under MSTAR dataset.
When the training samples each class are 20, the limited training SAR data have already provided the inner-class compactness as shown in Fig. \ref{ablation_fig1}(j), thus, the noise-invariance loss works better than the inner-class invariant proxy. This may illustrate that if the feature distribution of each class are compact enough, the inter-class separability is not that important.
By comparing four configurations in Fig. \ref{ablation_fig2}, even the training samples each class are 30, one individual invariance cannot improve the feature distributions of OpenSARShip dataset, though the generalization of V2 and V3 is improved. But the cooperation of dual invariance may still work, as showin in Fig. \ref{ablation_fig2}(d).

Based on the conducted ablation experiments and visualization of feature distributions under the limitations of limited SAR training samples, we have validated the effectiveness and soundness of our method. In the following sections, we will present the recognition performances of our method using different benchmark datasets.

\subsection{Recognition Results with Limited Training Samples under Different Datasets}

In this section, we present the recognition results of our method on the MSTAR, OpenSARship, and FUSAR-Ship datasets, taking into account the constraints of limited SAR training samples. We first describe the experimental configuration, followed by the presentation and analysis of the recognition results.

\subsubsection{Recognition Results under MSTAR}

We evaluate the recognition performance of our proposed method with limited SAR training samples on the MSTAR dataset, which consists of ten different targets. 
The experiments are conducted under the standard operating conditions (SOC) setup, where the training data is collected at a $\text{17}{}^\circ$ depression angle and the testing data is collected at a $\text{15}{}^\circ$ depression angle. 
The distribution of training and testing images in this setup is provided in Table \ref{ttnumMSTAR}. It is important to note that the numbers in the table represent the count of the original SAR images in the MSTAR dataset. 
In subsequent experiments, when referring to a "10-way n-shot" experiment, the value of $n$ represents the number of randomly selected images from the entire dataset as shown in Table \ref{ttnumMSTAR}.

Table \ref{mstar-results} presents the quantitative results of the recognition performance of our proposed method. 
The first row indicates the number of training images available for each target class. The targets are labeled in the first column according to their class and series. 
It is important to note that there are no additional training datasets or support datasets used, only a $n$-shot setup for each class. 
Following the common preprocesss \cite{my1}, the $n$ training images are augmented by a factor of 10 through random sampling of 10 image chips from a $384 \times 384$ SAR image, which is interpolated to $224 \times 224$ size to ensure the completeness of the central target. 
The recognition ratios for each of the ten targets in the MSTAR dataset, as well as the average recognition ratio, are calculated based on 20 experiments and presented in Table \ref{mstar-results}.

The data reveals that when each class has a minimum of 20 training samples, the recognition rates can surpass 96.00\%. If the number of training samples exceeds 60 per class, the recognition rates climb above 99.00\%. However, the situation changes significantly when the training samples per class are restricted to a range of 2-10. The recognition rates fluctuate noticeably in this case.
When training is limited to 5-shot and 2-shot scenarios, there are a total of 50 or 20 SAR images for training before augmentation. Despite these limitations, our proposed method manages to achieve recognition accuracies of 79.26\% and 58.93\% for ten-class recognition.
In a detailed examination of recognition rates for ten targets under a 2-shot scenario, it's observed that some targets are still identified well despite the reduction in training samples. However, as the number of training samples dwindles, the recognition performance for BMP2-9563, BTR60-7532, BRDM2-E7, T62-A51, and ZIL131-E12 is notably more impacted compared to the other five target types.

Based on the experimental results and analysis, our proposed method demonstrates excellent recognition performance even with a limited number of training samples ranging from 5 to 200 for each target type in the ten classes under the MSTAR dataset.

\subsubsection{Recognition Results of 3 and 6 classes under OpenSARship }

The OpenSARShip dataset consists of several ship classes that represent a significant portion of the international shipping market \cite{intro_aug_consit2}. In line with previous studies \cite{open2,open3,reduce1}, the 3-class experiment included bulk carriers, container ships, and tanks. The 6-class experiment in the OpenSARShip dataset consists of bulk carriers, container ships, tanks, cargo ships, fishing vessels, and general cargo ships.The preprocessing method for training and testing images followed the same procedure as the experiments conducted on the MSTAR dataset.

The recognition ratios presented in Table \ref{opensarship3-r} and Table \ref{opensarship6-r} were obtained by varying the number of training samples per class, ranging from 10 to 200. This range was selected based on previous studies in the field \cite{open2,open3,reduce1,intro_aug_consit2}.

The results show that our method demonstrates superior performance in recognizing SAR ship images across three classes. For the recognition of these three classes, our method exhibits robustness when the number of training samples per class ranges from 20 to 40.
Specifically, as the number of training samples in each class reduces from 40 to 20, there is only a 1.16\% drop in recognition rates, decreasing from 75.73\% to 74.57\%. When the training samples decrease from 80 to 40 in each class, the recognition rates drop by 4.88\%, moving from 80.61\% to 75.73\%. Additionally, when the training samples reduce from 200 to 100 in each class, the recognition rates drop by 2.50\%, going from 85.30\% to 82.80\%.
The same phenomenon is also shown in the recognition of 6 classes.

Thus, our method proves its robustness when confronted with reduced training samples, showing only marginal drops in recognition performance. This robustness in the face of limited training samples is a desirable attribute for SAR ATR methods, making them more viable for practical applications.

\subsubsection{Recognition Results under FUSAR-Ship}

As observed, the recognition task on the FUSAR-Ship dataset presents greater complexity compared to the OpenSARShip dataset. The FUSAR-Ship dataset includes five ship classes, which are among the most common in the global shipping market, and an additional class labeled as "other types of ships." The latter class encompasses a wide range of ship types beyond the first five classes, leading to increased overlap and requiring enhanced robustness and efficiency of the proposed method. The composition of the original training and testing sets in the FUSAR-Ship dataset is illustrated in Table \ref{fusarset}.

Table \ref{fusar-r} showcases the recognition performance of our method on the FUSAR-Ship dataset across five classes, with the number of training samples varying from 200 to 10. 
Based on the results, our method demonstrates superior recognition performance. When the number of training samples decreases from 200 to 60, there's a 5.28\% decrease in recognition rates, going from 68.67\% down to 63.39\%. Furthermore, as the training samples drop from 40 to 20, the recognition rates fall by 3.38\%, moving from 60.09\% to 57.71\%.
The more significant drop of 6.28\% between the recognition rates of 20 and 10 for each class could be attributed to the inherent complexity of the FUSAR-Ship dataset.

The performance of our method on the MSTAR, OpenSARship, and FUSAR-Ship datasets attests to its effectiveness and robustness, even when dealing with limited training samples. Our method has demonstrated its capacity to manage a range of imaging scenes, target classes (including both vehicles and ships), and complex imaging conditions. These results underscore the versatility and adaptability of our approach across diverse scenarios where training data is limited.

\renewcommand{\arraystretch}{2}
\begin{table}[htb!]
\centering
\footnotesize
\setlength\tabcolsep{5.3pt}
\caption{Comparison of Performances (\%) of 3 classes under OpenSARShip (The number in parentheses is the number of the training samples for each method)}
\label{3comparisonOPEN}
\begin{tabular}{c|c|c|c}
\toprule \toprule
\multirow{2}{*}{Methods} &
  \multicolumn{3}{c}{\begin{tabular}[c]{@{}c@{}}Bands of training \\ images each class\end{tabular}} \\ \cline{2-4} 
 &
 1 to 50 &
  51 to 100 &
  101 to 338 \\ \midrule
CNN\cite{compared2} &
  62.75 (50) &
  68.52 (100) &
  73.68 (200) \\ \hline
CNN+Matrix\cite{compared2} &
  72.86 (50) &
  75.31 (100) &
  77.22 (200) \\ \hline 
PFGFE-Net\cite{compared3} &
  - &
  - &
  79.84 (338) \\ \hline
MetaBoost\cite{compared4} &
  - &
  - &
  80.81 (338) \\ \hline
Semi-Supervised \cite{reduce1} &
  \begin{tabular}[c]{@{}c@{}}61.88 (20)\\ 64.73 (40)\end{tabular} &
  68.67 (80) &
  \begin{tabular}[c]{@{}c@{}}71.29 (120)\\ 74.96 (240)\end{tabular} \\ \hline
Supervised \cite{reduce1} &
  \begin{tabular}[c]{@{}c@{}}58.24 (20)\\ 62.09 (40)\end{tabular} &
  65.63 (80) &
  \begin{tabular}[c]{@{}c@{}}68.75 (120)\\ 70.83 (240)\end{tabular} \\ \hline
\begin{tabular}[c]{@{}c@{}}SM-CNN \cite{addopensarcp3}\\ (4096-4096-10)\end{tabular}  &
  - &
  - &
  81.80 (338) \\ \hline
KIDA \cite{addopensarcp4} &
  - &
  - &
  82.84 (338) \\ \hline
Proposed &
  \begin{tabular}[c]{@{}c@{}}74.57 (20)\\ 75.73 (40)\end{tabular} &
  \begin{tabular}[c]{@{}c@{}}80.61 (80)\\ 82.80 (100)\end{tabular} &
  85.30 (200) \\ \bottomrule \bottomrule
\end{tabular} 
\end{table}

\renewcommand{\arraystretch}{1.2}
\begin{table*}[htb!]
\centering
\footnotesize
\caption{Comparison with effective deep learning methods and specific methods for SAR ATR under constant training samples of OpenSARShip. (The training samples of methods under 3 classes and 6 classes are evaluated under 338 and 200 samples each class.)}
\label{3&6COMPARISONOpenSARShip}
\setlength\tabcolsep{1.5pt}
\begin{tabular}{c|c|cccccccc}
\toprule \toprule 
\multicolumn{2}{c|}{\multirow{2}{*}{Model}} & \multicolumn{4}{c}{3 classes}                                                                                                                                                                                                                                                 & \multicolumn{4}{c}{6 classes}                          \\ \cline{3-10} 
\multicolumn{2}{c|}{}    & Recall (\%)                                                       & Precision (\%)                                                    & F1 (\%)                                                           & Acc (\%)                                                          & Recall (\%) & Precision (\%) & F1 (\%)    & Acc (\%)   \\ \midrule
\multirow{15}{*}{\begin{tabular}[c]{@{}c@{}}Effective \\ deep learning \\ methods \end{tabular}} & LeNet-5 \cite{2compared1}                & 65.15±1.12 & 60.54±2.47 & 62.73±1.52 & 65.74±1.50 & 49.14±0.64 & 39.53±0.54 & 43.81±0.81 & 46.59±1.50 \\
& AlexNet \cite{2compared2}               & 68.51±3.04 & 65.52±1.23 & 66.94±1.51 & 70.22±0.68 & 53.40±0.61 & 44.39±0.60 & 48.48±0.87 & 50.98±2.14 \\
& VGG-11 \cite{2compared3} & 73.21±0.96 & 68.64±1.49 & 70.85±1.07 & 73.42±0.75 & 51.38±0.82 & 43.67±1.23 & 47.21±1.25 & 49.41±0.99 \\
& VGG-13 \cite{2compared3} & 72.59±1.29 & 67.24±1.75 & 68.79±0.92 & 73.03±0.86 & 51.32±0.38 & 43.06±1.68 & 46.83±0.90 & 49.70±1.36 \\
& GooLeNet \cite{2compared5} & 69.73±2.70 & 68.80±1.81 & 69.21±1.19 & 73.80±1.32 & 54.47±0.95 & 44.96±1.76 & 49.25±0.70 & 49.76±1.56 \\
& ResNet-18 \cite{2compared6} & 73.76±1.61 & 69.40±1.92 & 71.49±1.04 & 74.64±0.68 & 50.19±0.47 & 42.85±1.20 & 46.23±0.35 & 45.91±0.43 \\
& ResNet-34 \cite{2compared6} & 71.43±2.72 & 68.11±1.73 & 69.69±1.47 & 73.40±1.09 & 48.12±0.57 & 42.18±0.57 & 44.95±0.83 & 48.27±2.75 \\
& ResNet-50 \cite{2compared6} & 71.67±1.71 & 66.79±1.27 & 69.13±1.04 & 72.82±0.75 & 50.27±1.21 & 43.32±1.32 & 46.54±2.50 & 49.80±1.70 \\
& DenseNet-121 \cite{2compared10} & 72.55±3.88 & 69.56±2.17 & 70.93±1.60 & 74.65±0.68 & 55.51±1.30 & 46.52±1.48 & 50.62±0.74 & 53.49±1.47 \\
& DenseNet-161 \cite{2compared10} & 72.54±3.39 & 67.77±1.46 & 70.02±1.51 & 73.39±0.79 & 54.98±0.82 & 47.57±0.82 & 51.01±1.63 & 54.27±3.41 \\
& MobileNet-v3-Large \cite{2compared12}     & 65.12±2.53 & 60.75±1.72 & 62.84±1.73 & 66.13±0.92 & 49.95±0.58 & 42.14±0.62 & 45.71±0.65 & 46.60±2.61 \\
& MobileNet-v3-Small \cite{2compared12} & 67.23±1.59 & 61.85±1.69 & 64.42±1.41 & 66.71±0.87 & 48.28±0.75 & 40.75±0.73 & 44.20±0.57 & 44.41±1.10 \\
& SqueezeNet \cite{2compared13}    & 71.47±1.31 & 66.73±1.70 & 69.01±1.28 & 72.15±1.25 & 53.24±0.75 & 45.55±0.79 & 49.10±0.85 & 53.12±1.12 \\
& Inception-v4 \cite{2compared14}           & 69.26±3.16 & 67.43±2.39 & 68.28±1.97 & 72.44±0.70 & 54.92±0.69 & 46.46±0.49 & 50.34±1.31 & 54.55±3.52 \\
& Xception \cite{2compared15} & 71.56±3.00 & 68.60±1.67 & 70.00±1.29 & 73.74±0.86 & 52.21±0.94 & 44.03±1.15 & 47.77±1.11 & 49.56±1.47 \\ \midrule
\multirow{8}{*}{\begin{tabular}[c]{@{}c@{}}Methods for \\ SAR ATR \end{tabular}}  &  Wang et al. \cite{p4}            & 57.72±1.37 & 58.72±4.76 & 58.12±2.67 & 69.27±0.27 & 50.53±1.85 & 41.77±1.34 & 45.73±2.48 & 48.43±3.71 \\
& Hou et al. \cite{FUSAR}             & 69.33±2.00 & 69.44±2.42 & 66.76±1.64 & 67.41±1.13 & 48.76±0.79 & 41.22±0.74 & 44.67±1.21 & 47.44±2.01 \\
& Huang et al. \cite{2compared18}           & 74.74±1.60 & 69.56±2.38 & 72.04±1.60 & 74.98±1.46 & 54.09±0.81 & 47.58±1.66 & 50.63±1.79 & 54.78±2.08 \\
& Zhang et al. \cite{open3}           & 77.87±1.14 & 73.42±1.06 & 75.05±1.10 & 78.15±0.57 & 54.20±1.09 & 46.66±1.07 & 50.15±1.24 & 53.77±3.63 \\
& Zeng et al. \cite{OpenSARShip}            & 74.99±1.55 & 74.05±1.75 & 74.52±1,02 & 77.41±1.74 & 55.66±1.23 & 47.16±1.70 & 50.96±1.18 & 55.26±2.36 \\
& Xiong et al. \cite{2compared21}          & 73.87±1.16 & 71.50±3.00 & 72.67±2.04 & 75.44±2.68 & 53.57±0.33 & 45.74±0.82 & 49.35±0.69 & 54.93±2.61 \\
& SF-LPN-DPFF \cite{2compared22}           & 78.83±1.32 & 76.45±1.16 & 77.62±1.23 & 79.25±0.83 & 54.49±0.70 & 48.61±1.32 & 51.38±1.26 & 56.66±1.54 \\ \midrule
\multicolumn{2}{c|}{Ours}           & \begin{tabular}[c]{@{}c@{}}\bf{83.88±1.09}\end{tabular} & \begin{tabular}[c]{@{}c@{}}\bf{86.34±1.21}\end{tabular} & \begin{tabular}[c]{@{}c@{}}\bf{85.09±1.13}\end{tabular}  & \begin{tabular}[c]{@{}c@{}}\bf{85.30±0.59}\end{tabular} & \begin{tabular}[c]{@{}c@{}}\bf{66.70±0.52}\end{tabular}   & \begin{tabular}[c]{@{}c@{}}\bf{67.59±1.10}\end{tabular}         & \begin{tabular}[c]{@{}c@{}}\bf{67.14±0.67}\end{tabular}   & \begin{tabular}[c]{@{}c@{}}\bf{68.67±0.41}\end{tabular} \\ \bottomrule \bottomrule
\end{tabular} 
\end{table*}

\subsection{Comparison with Other Methods with Limited Training Samples}
In this subsection, we compare the performance of our method with other existing methods under two different ranges of limited training sample numbers. It is worth noting that the methods specifically designed for limited SAR data only utilize a limited number of training samples (referred to as $K$-shot) from the MSTAR dataset. If a particular method employs additional unlabeled images or utilizes other techniques in the context of limited SAR data, we provide further details regarding the specific usage of that method.

\begin{table*}[h]
\renewcommand{\arraystretch}{1.5}
\centering
\footnotesize
\caption{Comparison of Performances (\%) in MSTAR}
\label{comparisonMSTAR1}
\begin{tabular}{c|c|ccccccccc}
\toprule \toprule 
\multicolumn{2}{c|}{\multirow{2}{*}{Algorithms}} & \multicolumn{9}{c}{Image Number for Each Class}                                                         \\ \cline{3-11} 
\multicolumn{2}{c|}{}                            & \multicolumn{1}{c}{10}  &  \multicolumn{1}{c}{20} & \multicolumn{1}{c}{40} & \multicolumn{1}{c}{55} &\multicolumn{1}{c}{80} &\multicolumn{1}{c}{110} &\multicolumn{1}{c}{165} &\multicolumn{1}{c}{220} & \multicolumn{1}{c}{All data} \\ \midrule
\multirow{6}{*}{Traditional} &PCA+SVM \cite{comparison1}           &-    & 76.43             & 87.95    &   -      & 92.48        &- &- &-     & 94.32                 \\
&ADaboost \cite{comparison1}         &-      & 75.68             & 86.45     & -       & 91.45   &- &- &-          & 93.51                 \\
&DGM \cite{comparison1}          &-          & 81.11             & 88.14        & -    & 92.85    &- &- &-         & 96.07                 \\
&LNP \cite{addc1}  &-  &- & -  & 92.04  &   -      & 94.11  & 95.97      & 96.05         &  -        \\
&PSS-SVM \cite{addc2}     &-      &- &  -              & 95.01        & -  & 95.67           & 96.02          & 96.11  & -         \\ \midrule
\multirow{6}{*}{\begin{tabular}[c]{@{}c@{}}Data\\augmentation-based\end{tabular}} &GAN-CNN1 \cite{comparison1}       &-            & 81.80             & 88.35      & -      & 93.88    &- &- &-         & 97.03                 \\
& GAN-CNN2 \cite{comparison1}    &-            & 84.39             & 90.13    &   -      & 94.91   &- & -& -         & 97.53                 \\ 
&MGAN-CNN \cite{comparison1}      &-         & 85.23             & 90.82    & -        & 94.91   & -&- & -         & 97.81                 \\ 
&Triple-GAN\cite{addc3}   &-     &-&-               & 95.70      &-     & 95.97           & 96.13                   & 96.46      &-        \\
&Improved-GAN     \cite{addc4}   &-  &-&-             & 87.52        &-   & 95.02           & 97.26                     & 98.07    &-       \\
&Semi-supervised GAN\cite{addc5}   &-  &-&-            & 95.72      &-     & 97.22           & 97.97            & 98.14         &-    \\
&Supervised \cite{reduce1}  &-   &92.62   &97.11            &-      &98.65     &-           &-            &-         &-    \\ \midrule 
\multirow{11}{*}{\begin{tabular}[c]{@{}c@{}}Novel\\model/module\end{tabular}} &Deep CNN \cite{comparison2}         &-          & 77.86             & 86.98     &     -   & 93.04    &- &- &-         & 95.54                 \\
&Improved DNN \cite{addnew8}     &-              & 79.39             & 87.73    &  -       & 93.76    &- &- &  -       & 96.50                 \\
&Simple CNN \cite{compared2}      &-             & 75.88             & -     &      -      & -   &- &- &  -            & -                     \\
&Metric learning\cite{compared2}     &-        & 82.29             & -      &    -       & -      & -&- &    -       & -                     \\
&Dens-CapsNet\cite{ac2}      &80.26       & 92.95             & 96.50      &    -       & -      & -&- &    -       & 99.75                     \\
&LW-CMDANet\cite{ac3}      &-       & 55.34             & -      &    -       & -      & -&- &    -       & -                     \\ 
& FTL-dis\cite{ac5}      &81.21       & -             & -      &    -       & -      & -&- &    -       & -                     \\ 
& CL+pseudo-labels\cite{ac6}      &69.32       & -             & -      &    -       & -      & -&- &    -       & -                     \\ 
& HDLM\cite{ac9}      &88.16      & 95.17            & 97.85      &    -       & 98.80      & -&- &    -       & -                     \\ 
& ASC-MACN\cite{ac10}      &62.85      & 79.46            & -      &    -       & -      & -&- &    -       & 99.42                     \\ 
& ASC-MBCRN\cite{ac10}      &87.96      & 96.04            & -      &    -       & -      & -&- &    -       & 99.96                     \\
\midrule 
\multicolumn{2}{c|}{Ours   }                      & 91.18     & 96.45    & 98.64 & -  & 99.55  &- &- & - & -                     \\ \bottomrule \bottomrule
\end{tabular}
\end{table*}

\subsubsection{Comparison under OpenSARShip}
This subsection compares the performances of the proposed method with other state-of-the-art methods. Two types of comparisons were selected: one is the comparison with other SAR ship target recognition methods under the decreasing training data, and the other is the comparison with other effective deep learning networks under constant training sample number.

Comparison under decreasing training data. Table \ref{3comparisonOPEN} displays a comparison of various methods used for SAR ship recognition, which include semi-supervised learning, supervised learning, CNN, CNN + matrix, PFGFE-Net, and MetaBoost. 
The Supervised Learning \cite{reduce1} is a supervised variant of the model discussed in \cite{reduce1}. 
CNNs \cite{compared2} are common frameworks employed for classification tasks. CNN + Matrix \cite{compared2} integrates CNN and a matrix model to achieve superior performance. 
PFGFE-Net \cite{compared3} effectively fuses polarization information at the input data level, feature level, and decision level, thereby addressing the issue of insufficient utilization of polarization information.
MetaBoost \cite{compared4} applies a two-stage filtration process and primarily focuses on the generation and combination of "good and different" base classifiers. 
Three bands (from 1 to 50, from 51 to 100, and from 101 to 338) are defined based on the number of training images available for each class.

The performance comparison of three classes of training data on the OpenSARShip dataset is shown in Table \ref{3comparisonOPEN}.
Our method outperforms others in classifying three classes of training data on the OpenSARShip dataset, reaching recognition rates of 74.57\% and 75.73\% with 20 and 40 samples per class, respectively. In contrast, supervised learning achieves 58.24\% with 20 samples, and with 40 or 50 samples, recognition rates for supervised learning, CNN, and CNN + matrix don't surpass 72.86\%. This comparison clearly shows the superior performance of our method over other leading methods when training data is limited.

Our method exhibits exceptional performance across different bands of training data. With 80 and 100 samples per class, we achieve recognition rates of 80.61\% and 82.80\% respectively. In contrast, supervised learning only reaches 65.63\% with 80 samples, and CNN methods hover around 68.52\%-75.31\% with 100 samples. When samples increase to 200 per class, our method excels with an 85.30\% recognition rate, while others still lag behind, even with 240 samples. With 338 samples per class, state-of-the-art methods like PFGFE-Net and MetaBoost achieve rates below 81\%, demonstrating our method's superior performance despite limited training data.

Comparison under constant training data.
The quantitative comparison of our method with other modern deep learning-based approaches is illustrated in Table \ref{3&6COMPARISONOpenSARShip}. In this evaluation, we use four metrics - recall, precision, F1-score, and accuracy - to provide a comprehensive assessment of the model's performance. 
When performing recognition tasks under three classes, the proposed method demonstrates superior results with a recall of 83.88\%, precision of 86.34\%, F1-score of 85.09\%, and an overall accuracy of 85.30\%. These values outperform the best results from all other methods, which achieved a recall of 78.83\%, precision of 76.45\%, F1-score of 77.62\%, and accuracy of 79.25
In the context of six classes, the proposed method again outperforms the rest with a recall of 66.70\%, precision of 67.59\%, F1-score of 67.14\%, and accuracy of 68.67\%. The highest performing results from all other methods under the same conditions reached a recall of 54.49\%, precision of 48.61\%, F1-score of 51.38\%, and accuracy of 56.66\%.
These results validate the superiority of the proposed method, showcasing its effectiveness and robustness in performing SAR ship recognition tasks under varying numbers of classes and limited training data. The proposed method's consistently high performance indicates its potential in practical applications of SAR ATR.

The aforementioned comparisons clearly demonstrate that our proposed method can achieve leading-edge recognition performance across a broad range of training samples. Furthermore, our method distinctly surpasses all other techniques under two types of recognition involving varying class numbers.

\subsubsection{Comparison under MSTAR}

To assess the impact of limited SAR training samples, we compare our method with state-of-the-art approaches designed for limited SAR data in Table \ref{comparisonMSTAR}. The comparison is performed across different sample size scenarios, ranging from the full dataset to a reduced number of 20 training samples.
MGAN-CNN \cite{comparison1} utilizes a multi-discriminator architecture to improve the generated image quality of Generative Adversarial Networks (GANs), thereby enhancing recognition performance. CNN1 and GAN-CNN are simplified versions of MGAN-CNN.
The Semisupervised method \cite{reduce1} proposes a self-consistent augmentation technique to leverage unlabeled data. For fair comparison, we only consider the recognition results obtained without utilizing sufficient unlabeled images in this paper.
In addition to the aforementioned methods, we also include other well-known approaches for comparison, such as PCA+SVM, ADaboost, LC-KSVD, DGM, and various DNN-based and CNN-based methods (DNN1, DNN2, CNN2, and CNN+matrix) \cite{reduce1}.

By comparing the recognition performances of these methods with our method, we can gain insights into the effectiveness of our method in the context of limited SAR training samples.
The proposed method outperforms other techniques in terms of recognition performance, especially when the number of training images per class is small. It maintains high accuracy even with limited training data, demonstrating robustness and adaptability in data-constrained scenarios.

The performance comparison of our method against other limited SAR data techniques on the MSTAR dataset clearly shows that our approach has achieved top-tier results. Even in the face of limited training samples and a significant disparity between training and testing samples, our method outperforms its counterparts.

Moreover, the recognition comparisons under the constraints of limited SAR training samples on both the MSTAR and OpenSARship datasets affirm the efficacy and applicability of our approach. Our method sets a benchmark with only a handful of training samples, even under varying imaging conditions across different datasets. Notably, our method exhibits resilience when dealing with extremely few training samples and a substantial variation between training and testing samples, which bolsters the practical implementation of SAR ATR methods.

\section{Conclusion}\label{conclusions}
This paper has investigated the critical problem of weak generalization in SAR ATR due to limited data availability. 
By constructing a causal ATR model, we elucidated the role of noise as a confounder, impairing the efficacy of feature extraction from SAR images in limited data situations.
While the detrimental impact of noise on feature extraction can theoretically be estimated and eliminated via backdoor adjustment, it remains challenging due to inter-class variability and intrinsic-similarity confusion, and the limitations of ERM with limited SAR data. 
In addressing these sub-problems, the proposed dual invariance, encompassing an inner-class invariant proxy and noise-invariance loss, has proven effective. 
The inner-class invariant proxy assists in accurately estimating the effect of noise, while the noise-invariance loss successfully mitigates the noise impact, transforming the data quantity requirement of ERM into a necessity for noise environment annotations.
The proposed model successfully uncovers the core problem in SAR ATR with limited data and provides a robust solution. 
Experimental results, comparative analysis, and ablation studies on the MSTAR, OpenSARship, and FUSAR-Ship datasets confirm the superior performance of our proposed method in terms of recognition. 
We believe this research contributes to a more accurate understanding of the challenges posed by limited SAR data, and providing a practical solution for mitigating the impacts of noise on feature.





\section*{Acknowledgments}\label{ACKNOWLEDGEMENTS}
This work was supported by the China Scholarship Council.
Thanks to the reviewers and editors for their efforts and assistance.

\appendix
\section{Related Works}
\label{Appendix1}
\subsection{SAR ATR with Limited Data} 
Recent years have seen a growing body of research tackling the issue of insufficient SAR images in SAR ATR. 
In scenarios characterized by a paucity of SAR images, there exist one training dataset and one testing dataset. Each target class type in the training dataset has a limited number of samples. For instance, if there are 10 SAR images for all 10 class types, the training dataset contains $10 \times 10 = 100$ SAR images in total. The models are trained and tested using only these 100 SAR images. At times, some methodologies construct a semi-supervised structure to utilize the remaining images as unlabeled samples, in addition to the $K$ labeled images during training. Under these circumstances, the input information supplied to the model exceeds the $K$ labeled images.

For example, Wang et al. \cite{reduce1} devised a semi-supervised learning framework that chiefly includes a self-consistent augmentation rule to make use of unlabeled data during training for limited SAR images. Zhang et al. \cite{fsldata4} employed feature augmentation and ensemble learning strategies to concatenate cascaded features from optimally selected convolutional layers, thereby extracting a more comprehensive representation of information from limited data. Sun et al. \cite{fsldata3} proposed an attribute-guided transfer learning method that uses an angular rotation generative network. The shared attribute between the source and target domains in this method is the target aspect angle, which addresses the problem of a lack of training data at different aspect angles. Zhang et al. \cite{readd1} designed a semi-supervised transfer learning method for limited SAR training data, using learned parameters from a pre-trained GAN, achieving up to a 23.58\% accuracy improvement compared with other random-initialized models. Moreover, they suggested another transfer learning method for limited SAR training data, in which pre-trained layers are reused to transfer generic knowledge \cite{readd2}.

\subsection{Causal Theory}
Causality has emerged as a crucial concept in deep learning, aiming to transform the field from a mostly correlative to a more causal discipline. Here, we review some of the key work related to causality and its application in deep learning.
Causal Inference in Deep Learning Models. Early efforts for integrating causal principles in deep learning focused on enabling neural networks to understand and model the cause-effect relationships \cite{ramachandra2018deep, luo2020causal, cui2020causal}. 
Causal representation learning. Representation learning is a central aspect of deep learning. Recently, there's been increasing interest in learning representations that capture causal structures \cite{wang2021desiderata,mitrovic2020representation, scholkopf2021toward} . 
Counterfactual Reasoning in Deep Learning. Counterfactual reasoning, a central concept in causal inference, has found its application in deep learning \cite{oh2022learn, liu2019generative,pfohl2019counterfactual}. 
Causal Reinforcement Learning. The application of causality in reinforcement learning has been of particular interest, with an aim to learn policies that generalize across different environments \cite{zhu2019causal,madumal2020explainable, dasgupta2019causal}. 
These works form the basis of our understanding of causality's role in deep learning, and the field continues to develop at a rapid pace. The application of causality in deep learning not only opens up new methodologies but also contributes to a more fundamental understanding of learning algorithms. 

Although causality has made many advancements in deep learning, it has not yet shown its guiding role and high interpretability in the field of Synthetic Aperture Radar Automatic Target Recognition (SAR ATR).

\section{Causal Perspective between Ample and Limited SAR Data}
\label{Appendix2}
\subsection{Casual inference fundamentals}
\label{Appendix21}
In the context of causal inference using Directed Acyclic Graphs (DAGs), there are three fundamental structures \cite{glymour2016causal}: Chain, Fork, and Collider:

1. \textbf{Chain:} This structure represents a sequential relationship where node ($M$) acts as a mediator between two nodes $(X \text{ and } Y)$. The visual representation of this structure is $X \rightarrow M \rightarrow Y$. $X$ and $Y$ are dependent due to the shared influence of $M$. However, when conditioning on $M$ (i.e., holding the value of $M$ constant or controlled), $X$ and $Y$ become independent and then the causal relation between $X$ and $Y$ is broken.

2. \textbf{Fork:} This arrangement involves node ($M$) that gives rise to or precedes two other nodes ($X$ and $Y$). It can be visually depicted as $X \leftarrow M \rightarrow Y$. In this structure, $X$ and $Y$ are dependent due to the common cause $M$. When conditioning on $M$, $X$ and $Y$ become independent as the common source of their correlation is taken into account.

3. \textbf{Collider:} This structure occurs when node ($M$) is affected by or is a result of two other nodes ($X$ and $Y$). It is visually represented as $X \rightarrow  M \leftarrow  Y$. In a collider, $X$ and $Y$ are initially independent, but conditioning on $M$ leads to their dependence. This dependence arises because conditioning on M opens a path of influence between $X$ and $Y$, often referred to as "collider bias" or "collider-stratification bias".

After establishing these basic structures, the concept of d-separation is further introduced as a critical criterion used in SCMs to examine the dependencies between nodes (data variables) \cite{pearl2009causality}.

\textbf{d-separation:} In a given causal DAG $\mathcal{G}$, a set of nodes $Z$ is said to d-separate two other nodes $X$ and $Y$, if $Z$ manages to block all paths from $X$ to $Y$. A path is blocked by $Z$ if and only if: 1) $D$ which is within the set $Z$ is the middle variable between a \textbf{chain} or a \textbf{fork} e.g. $X\rightarrow D\rightarrow Y$, $X\leftarrow D\rightarrow Y$. 2) the path contains a collider $X\rightarrow D\leftarrow Y$ where D along with its descendants are not within set $Z$. In other words, if $X$ and $Y$ are d-separated given $Z$, then $X$ and $Y$ are conditionally independent given $Z$.

Lastly, another fundamental concept that we cover is the Instrumental Variable.

\textbf{Instrumental Variable:} An instrumental variable (IV) is a variable that is used to identify and estimate the causal effect of a particular variable of interest (the treatment variable) $X$ on the outcome $Y$ (i.e. $X\rightarrow Y$). A valid IV Z should be associated with the treatment variable (i.e. $X \not\perp Z$ in $\mathcal{G}$), but not have any direct association with the outcome $Y$ or any unmeasured confounding variables (i.e. $Z \perp Y$ in $\mathcal{G}$ manipulated where all the arrows going into node X are removed). The utility of IVs comes from their capacity to provide unbiased causal estimates even when unmeasured confounding is present.

\subsection{Causal Difference between Ample and
Limited SAR Data}

In this section, we will delve into the causal difference of SAR ATR between ample and limited SAR data.

\textbf{Ample SAR Data}: In instances where the volume of SAR data is ample, the probability of possessing more target attributes, denoted as $A$, escalates compared to situations with limited SAR data. When analyzing the $i$-th SAR image, which encompasses a subset of $A$, and its extracted features represented as $X$, the ATR model confronts a daunting task of developing a direct one-to-one correlation between $X$ and $i$. This complexity emerges from two key factors:
1) In the context of ample SAR data, the process of establishing a one-to-one correlation with the ATR model can be compared to finding a needle in a haystack.
2) The significant magnitude of SAR data boosts the chance of encountering SAR images that present similar subsets of $A$.
Therefore, the causal relationship between $A$ and $X$ is unidirectional in nature, explicitly $A \to X$ with ample SAR data.

As a result, $A$ plays the role of an instrumental variable in the $A \to X \rightarrow N$ pathway, effectively creating a collider. This structural phenomenon ensures $A$ and $N$ remain independent even though they are connected via $X$, as mentioned in \ref{Appendix21}.
Hence, the ATR model, by incorporating $P(Y|X):=P(Y|A)$, guarantees that $N$ ceases to influence $Y$— that is, $P(Y|X) \approx P(Y|do(X))$, thereby accomplishing accurate recognition performance along with robust generalization as portrayed in Fig. \ref{causal}(a). It is worth highlighting that $P(Y|do(X))$ signifies the ideal ATR approach which effectively interrupts the backdoor pathway $N \to X$.

\textbf{Limited SAR Data}: On the other hand, with scarce SAR data, the pathway $X \to A$ is established owing to the relative simplicity of the ATR method to create a one-to-one correlation between $X$ and $i$, demonstrated in Fig. \ref{causal}(b). Under such circumstances, $P(Y|X) \not\approx P(Y|do(X))$ via the pathway $X \rightarrow N \to Y$. 
In effect, within the scope of limited SAR data, the noise $N$ incites illusory correlations in the ATR method, thereby undermining its generalization potential. 

The causal solution of weak generalization of ATR with limited SAR data is to chase the direct causality between $X$ and $Y$ just via $X \to Y$ using backdoor adjustment, as shown in Fig. \ref{causal}(c). The derivation of the proposed causal model using backdoor adjustment is shown in next section.

\section{The Derivation of Backdoor Adjustment For The Proposed Causal Model}
\label{Appendix3}
In this section, we will formally derive the backdoor adjustment for the proposed causal graph in Fig. \ref{causal}, using $do$-calculus devised by Judea Pearl \cite{pearl2012calculus}. We have the set of three formal rules of the $do$-calculus below:\\
    \textbf{Rule 1:} Insertion/Deletion of Observations\par
    If $Y\perp X|Z$ in $\mathcal{G}$, then:
    \begin{equation}
        P(Y | do(X), Z, W) = P(Y | Z, W)
    \end{equation}
    Here, W denotes a set of variables that doesn't include X or Y.\\
    \textbf{Rule 2:} Action/Observation Exchange\par
    If the distribution of $X$ is not affected by any of the variables in $Z$, then:
    \begin{equation}
        P(Y | do(X), do(Z), W) = P(Y | do(X), Z, W)
    \end{equation}
    In this rule, $W$ is a set of variables not including $X$, $Y$ or any of the variables in $Z$.\\
    \textbf{Rule 3:} Insertion/Deletion of Actions\par
    If $X\perp Z|W$ in $\mathcal{G}$ obtained by removing all arrows into $X$, then:
    \begin{equation}
        P(Y | do(X), do(Z), W) = P(Y | do(X), W)
    \end{equation}
    In this rule, $W$ is a set of variables not including $X$, $Y$ or any of the variables in $Z$.\\

    In our causal graph, the target distribution under intervention $P(Y | do(X))$, can be derived as follows:
    \begin{align}
        P(Y|do(X)) &= \sum_N P(Y|do(X), N)P(N|do(X)) \label{B.4} \\
        &= \sum_N P(Y|do(X), N)P(N) \label{B.5} \\
        &= \sum_N P(Y|X, N)P(N) \label{B.6}
 \end{align}
 
In the above equation, we first apply the law of total probability to obtain Eq.(\ref{B.4}).  Eq.(\ref{B.5}) employs Rule 3 under the condition that $N \perp X$ in $\mathcal{G}$ removing all arrows into $X$. Eq.(\ref{B.6})  applies Rule 2 to transform the intervention term to an observation term, denoted as $(Y  \perp X|D)$ in $\mathcal{G}$ removing all arrows outgoing $X$.

\bibliographystyle{elsarticle-num} 
\bibliography{bib/references, bib/mypub, bib/causal_IRM}

\begin{thebibliography}{100}
\expandafter\ifx\csname url\endcsname\relax
  \def\url#1{\texttt{#1}}\fi
\expandafter\ifx\csname urlprefix\endcsname\relax\def\urlprefix{URL }\fi
\expandafter\ifx\csname href\endcsname\relax
  \def\href#1#2{#2} \def\path#1{#1}\fi

\bibitem{intro1}
J.~C. Curlander, R.~N. McDonough, Synthetic aperture radar, Vol.~11, Wiley, New
  York, 1991.

\bibitem{intro2}
B.~Bhanu, Automatic target recognition: State of the art survey, IEEE
  transactions on aerospace and electronic systems~(4) (1986) 364--379.

\bibitem{addnew1}
K.~El-Darymli, E.~W. Gill, P.~Mcguire, D.~Power, C.~Moloney, Automatic target
  recognition in synthetic aperture radar imagery: A state-of-the-art review,
  IEEE access 4 (2016) 6014--6058.

\bibitem{reff1}
X.~Yan, M.~Jia, A novel optimized svm classification algorithm with
  multi-domain feature and its application to fault diagnosis of rolling
  bearing, Neurocomputing 313 (2018) 47--64.

\bibitem{reff2}
E.~Syriani, L.~Luhunu, H.~Sahraoui, Systematic mapping study of template-based
  code generation, Computer Languages, Systems \& Structures 52 (2018) 43--62.

\bibitem{reff3}
G.~Orr{\`u}, G.~L. Marcialis, F.~Roli, A novel classification-selection
  approach for the self updating of template-based face recognition systems,
  Pattern Recognition 100 (2020) 107121.

\bibitem{my1}
C.~Wang, J.~Pei, Z.~Wang, Y.~Huang, J.~Wu, H.~Yang, J.~Yang, When deep learning
  meets multi-task learning in sar atr: Simultaneous target recognition and
  segmentation, Remote Sensing 12~(23) (2020) 3863.

\bibitem{addnew2}
G.~J. Owirka, S.~M. Verbout, L.~M. Novak, Template-based sar atr performance
  using different image enhancement techniques, in: Algorithms for Synthetic
  Aperture Radar Imagery VI, Vol. 3721, SPIE, 1999, pp. 302--319.

\bibitem{addnew3}
L.~M. Novak, G.~J. Owirka, W.~S. Brower, A.~L. Weaver, The automatic
  target-recognition system in saip, Lincoln Laboratory Journal 10~(2) (1997).

\bibitem{addnew4}
Q.~Zhao, J.~C. Principe, Support vector machines for sar automatic target
  recognition, IEEE Transactions on Aerospace and Electronic Systems 37~(2)
  (2001) 643--654.

\bibitem{addnew5}
Y.~Sun, Z.~Liu, S.~Todorovic, J.~Li, Adaptive boosting for sar automatic target
  recognition, IEEE Transactions on Aerospace and Electronic Systems 43~(1)
  (2007) 112--125.

\bibitem{addnew6}
J.~C. Principe, A.~Radisavljevic, J.~Fisher, M.~Hiett, L.~M. Novak, Target
  prescreening based on a quadratic gamma discriminator, IEEE Transactions on
  Aerospace and Electronic Systems 34~(3) (1998) 706--715.

\bibitem{ATR3}
R.~Xue, X.~Bai, F.~Zhou, Spatial--temporal ensemble convolution for sequence
  sar target classification, IEEE Transactions on Geoscience and Remote Sensing
  59~(2) (2020) 1250--1262.

\bibitem{isprs3}
S.~{Temitope Yekeen}, A.~Balogun, K.~B. {Wan Yusof},
  \href{https://www.sciencedirect.com/science/article/pii/S0924271620301982}{A
  novel deep learning instance segmentation model for automated marine oil
  spill detection}, ISPRS Journal of Photogrammetry and Remote Sensing 167
  (2020) 190--200.
\newblock \href
  {https://doi.org/https://doi.org/10.1016/j.isprsjprs.2020.07.011}
  {\path{doi:https://doi.org/10.1016/j.isprsjprs.2020.07.011}}.
\newline\urlprefix\url{https://www.sciencedirect.com/science/article/pii/S0924271620301982}

\bibitem{ATR1}
S.~Chen, H.~Wang, F.~Xu, Y.-Q. Jin, Target classification using the deep
  convolutional networks for sar images, IEEE transactions on geoscience and
  remote sensing 54~(8) (2016) 4806--4817.

\bibitem{isprs2}
J.~Chen, X.~Qiu, C.~Ding, Y.~Wu,
  \href{https://www.sciencedirect.com/science/article/pii/S0924271622000946}{Sar
  image classification based on spiking neural network through spike-time
  dependent plasticity and gradient descent}, ISPRS Journal of Photogrammetry
  and Remote Sensing 188 (2022) 109--124.
\newblock \href
  {https://doi.org/https://doi.org/10.1016/j.isprsjprs.2022.03.021}
  {\path{doi:https://doi.org/10.1016/j.isprsjprs.2022.03.021}}.
\newline\urlprefix\url{https://www.sciencedirect.com/science/article/pii/S0924271622000946}

\bibitem{isprs6}
T.~Zhang, X.~Zhang, J.~Shi, S.~Wei,
  \href{https://www.sciencedirect.com/science/article/pii/S0924271620301404}{Hyperli-net:
  A hyper-light deep learning network for high-accurate and high-speed ship
  detection from synthetic aperture radar imagery}, ISPRS Journal of
  Photogrammetry and Remote Sensing 167 (2020) 123--153.
\newblock \href
  {https://doi.org/https://doi.org/10.1016/j.isprsjprs.2020.05.016}
  {\path{doi:https://doi.org/10.1016/j.isprsjprs.2020.05.016}}.
\newline\urlprefix\url{https://www.sciencedirect.com/science/article/pii/S0924271620301404}

\bibitem{addnew7}
K.~Du, Y.~Deng, R.~Wang, T.~Zhao, N.~Li, Sar atr based on displacement-and
  rotation-insensitive cnn, Remote Sensing Letters 7~(9) (2016) 895--904.

\bibitem{li2023panoptic}
L.~Li, W.~Ji, Y.~Wu, M.~Li, Y.~Qin, L.~Wei, R.~Zimmermann, Panoptic scene graph
  generation with semantics-prototype learning (2023).
\newblock \href {http://arxiv.org/abs/2307.15567} {\path{arXiv:2307.15567}}.

\bibitem{liang2023efficient}
R.~Liang, Y.~Yang, H.~Lu, L.~Li, Efficient temporal sentence grounding in
  videos with multi-teacher knowledge distillation (2023).
\newblock \href {http://arxiv.org/abs/2308.03725} {\path{arXiv:2308.03725}}.

\bibitem{addnew8}
J.~Ding, B.~Chen, H.~Liu, M.~Huang, Convolutional neural network with data
  augmentation for sar target recognition, IEEE Geoscience and remote sensing
  letters 13~(3) (2016) 364--368.

\bibitem{addnew9}
N.~Wang, Y.~Wang, H.~Liu, Q.~Zuo, J.~He, Feature-fused sar target
  discrimination using multiple convolutional neural networks, IEEE Geoscience
  and remote sensing letters 14~(10) (2017) 1695--1699.

\bibitem{reff4}
S.~Li, W.~Song, L.~Fang, Y.~Chen, P.~Ghamisi, J.~A. Benediktsson, Deep learning
  for hyperspectral image classification: An overview, IEEE Transactions on
  Geoscience and Remote Sensing 57~(9) (2019) 6690--6709.

\bibitem{reff5}
Y.~Li, H.~Zhang, X.~Xue, Y.~Jiang, Q.~Shen, Deep learning for remote sensing
  image classification: A survey, Wiley Interdisciplinary Reviews: Data Mining
  and Knowledge Discovery 8~(6) (2018) e1264.

\bibitem{reff6}
A.~Miko{\l}ajczyk, M.~Grochowski, Data augmentation for improving deep learning
  in image classification problem, in: 2018 international interdisciplinary PhD
  workshop (IIPhDW), IEEE, 2018, pp. 117--122.

\bibitem{addnew10}
M.~Touafria, Q.~Yang, A concurrent and hierarchy target learning architecture
  for classification in sar application, Sensors 18~(10) (2018) 3218.

\bibitem{reff7}
S.~Chen, H.~Wang, Sar target recognition based on deep learning, in: 2014
  International Conference on Data Science and Advanced Analytics (DSAA), IEEE,
  2014, pp. 541--547.

\bibitem{reff8}
S.~A. Wagner, Sar atr by a combination of convolutional neural network and
  support vector machines, IEEE transactions on Aerospace and Electronic
  Systems 52~(6) (2016) 2861--2872.

\bibitem{isprs1}
J.~Geng, W.~Jiang, X.~Deng,
  \href{https://www.sciencedirect.com/science/article/pii/S0924271620301945}{Multi-scale
  deep feature learning network with bilateral filtering for sar image
  classification}, ISPRS Journal of Photogrammetry and Remote Sensing 167
  (2020) 201--213.
\newblock \href
  {https://doi.org/https://doi.org/10.1016/j.isprsjprs.2020.07.007}
  {\path{doi:https://doi.org/10.1016/j.isprsjprs.2020.07.007}}.
\newline\urlprefix\url{https://www.sciencedirect.com/science/article/pii/S0924271620301945}

\bibitem{add1}
X.~Wang, Z.~Cao, Y.~Pi, Semisupervised classification with adaptive anchor
  graph for polsar images, IEEE Geoscience and Remote Sensing Letters 19 (2021)
  1--5.

\bibitem{add2}
Z.~Yue, F.~Gao, Q.~Xiong, J.~Wang, T.~Huang, E.~Yang, H.~Zhou, A novel
  semi-supervised convolutional neural network method for synthetic aperture
  radar image recognition, Cognitive Computation 13~(4) (2021) 795--806.

\bibitem{add3}
C.~Cao, Z.~Cao, Z.~Cui, Ldgan: A synthetic aperture radar image generation
  method for automatic target recognition, IEEE Transactions on Geoscience and
  Remote Sensing 58~(5) (2019) 3495--3508.

\bibitem{isprs5}
T.~Zhang, X.~Zhang, C.~Liu, J.~Shi, S.~Wei, I.~Ahmad, X.~Zhan, Y.~Zhou, D.~Pan,
  J.~Li, H.~Su,
  \href{https://www.sciencedirect.com/science/article/pii/S0924271621002781}{Balance
  learning for ship detection from synthetic aperture radar remote sensing
  imagery}, ISPRS Journal of Photogrammetry and Remote Sensing 182 (2021)
  190--207.
\newblock \href
  {https://doi.org/https://doi.org/10.1016/j.isprsjprs.2021.10.010}
  {\path{doi:https://doi.org/10.1016/j.isprsjprs.2021.10.010}}.
\newline\urlprefix\url{https://www.sciencedirect.com/science/article/pii/S0924271621002781}

\bibitem{r1}
Z.~Huang, X.~Yao, Y.~Liu, C.~O. Dumitru, M.~Datcu, J.~Han,
  \href{https://www.sciencedirect.com/science/article/pii/S0924271622001472}{Physically
  explainable cnn for sar image classification}, ISPRS Journal of
  Photogrammetry and Remote Sensing 190 (2022) 25--37.
\newblock \href
  {https://doi.org/https://doi.org/10.1016/j.isprsjprs.2022.05.008}
  {\path{doi:https://doi.org/10.1016/j.isprsjprs.2022.05.008}}.
\newline\urlprefix\url{https://www.sciencedirect.com/science/article/pii/S0924271622001472}

\bibitem{r2}
Z.~Huang, Z.~Pan, B.~Lei, What, where, and how to transfer in sar target
  recognition based on deep cnns, IEEE Transactions on Geoscience and Remote
  Sensing 58~(4) (2019) 2324--2336.

\bibitem{r3}
Z.~Wen, Z.~Liu, S.~Zhang, Q.~Pan, Rotation awareness based self-supervised
  learning for sar target recognition with limited training samples, IEEE
  Transactions on Image Processing 30 (2021) 7266--7279.
\newblock \href {https://doi.org/10.1109/TIP.2021.3104179}
  {\path{doi:10.1109/TIP.2021.3104179}}.

\bibitem{r4}
B.~Ren, Y.~Zhao, B.~Hou, J.~Chanussot, L.~Jiao, A mutual information-based
  self-supervised learning model for polsar land cover classification, IEEE
  Transactions on Geoscience and Remote Sensing 59~(11) (2021) 9224--9237.
\newblock \href {https://doi.org/10.1109/TGRS.2020.3048967}
  {\path{doi:10.1109/TGRS.2020.3048967}}.

\bibitem{r5}
J.~Zhang, M.~Xing, Y.~Xie, Fec: A feature fusion framework for sar target
  recognition based on electromagnetic scattering features and deep cnn
  features, IEEE Transactions on Geoscience and Remote Sensing 59~(3) (2020)
  2174--2187.

\bibitem{r6}
X.~Shi, F.~Zhou, S.~Yang, Z.~Zhang, T.~Su, Automatic target recognition for
  synthetic aperture radar images based on super-resolution generative
  adversarial network and deep convolutional neural network, Remote Sensing
  11~(2) (2019) 135.

\bibitem{r7}
Z.~Lin, K.~Ji, M.~Kang, X.~Leng, H.~Zou, Deep convolutional highway unit
  network for sar target classification with limited labeled training data,
  IEEE Geoscience and Remote Sensing Letters 14~(7) (2017) 1091--1095.

\bibitem{r8}
F.~Zhang, C.~Hu, Q.~Yin, W.~Li, H.-C. Li, W.~Hong, Multi-aspect-aware
  bidirectional lstm networks for synthetic aperture radar target recognition,
  IEEE Access 5 (2017) 26880--26891.

\bibitem{r9}
J.~H. Cho, C.~G. Park, Multiple feature aggregation using convolutional neural
  networks for sar image-based automatic target recognition, IEEE Geoscience
  and Remote Sensing Letters 15~(12) (2018) 1882--1886.

\bibitem{r10}
Q.~Yu, H.~Hu, X.~Geng, Y.~Jiang, J.~An, High-performance sar automatic target
  recognition under limited data condition based on a deep feature fusion
  network, IEEE Access 7 (2019) 165646--165658.

\bibitem{pearl2009causality}
J.~Pearl, Causality, Cambridge university press, 2009.

\bibitem{morgan2015counterfactuals}
S.~L. Morgan, C.~Winship, Counterfactuals and causal inference, Cambridge
  University Press, 2015.

\bibitem{pearl2009causal}
J.~Pearl, Causal inference in statistics: An overview (2009).

\bibitem{glymour2016causal}
M.~Glymour, J.~Pearl, N.~P. Jewell, Causal inference in statistics: A primer,
  John Wiley \& Sons, 2016.

\bibitem{pearl2012calculus}
J.~Pearl, The do-calculus revisited, arXiv preprint arXiv:1210.4852 (2012).

\bibitem{pfister2019invariant}
N.~Pfister, P.~B{\"u}hlmann, J.~Peters, Invariant causal prediction for
  sequential data, Journal of the American Statistical Association 114~(527)
  (2019) 1264--1276.

\bibitem{scholkopf2012causal}
B.~Sch{\"o}lkopf, D.~Janzing, J.~Peters, E.~Sgouritsa, K.~Zhang, J.~Mooij, On
  causal and anticausal learning, arXiv preprint arXiv:1206.6471 (2012).

\bibitem{greenland1999causal}
S.~Greenland, J.~Pearl, J.~M. Robins, Causal diagrams for epidemiologic
  research, Epidemiology (1999) 37--48.

\bibitem{IRM}
M.~Arjovsky, L.~Bottou, I.~Gulrajani, D.~Lopez-Paz, Invariant risk
  minimization, arXiv preprint arXiv:1907.02893 (2019).

\bibitem{RIRM}
E.~Rosenfeld, P.~Ravikumar, A.~Risteski, The risks of invariant risk
  minimization, arXiv preprint arXiv:2010.05761 (2020).

\bibitem{fslmodel4}
M.~Yang, X.~Bai, L.~Wang, F.~Zhou, Mixed loss graph attention network for
  few-shot sar target classification, IEEE Transactions on Geoscience and
  Remote Sensing 60 (2022) 1--13.
\newblock \href {https://doi.org/10.1109/TGRS.2021.3124336}
  {\path{doi:10.1109/TGRS.2021.3124336}}.

\bibitem{open2}
T.~Zhang, X.~Zhang, X.~Ke, C.~Liu, X.~Xu, X.~Zhan, C.~Wang, I.~Ahmad, Y.~Zhou,
  D.~Pan, et~al., Hog-shipclsnet: A novel deep learning network with hog
  feature fusion for sar ship classification, IEEE Transactions on Geoscience
  and Remote Sensing 60 (2021) 1--22.

\bibitem{fslmodel2}
L.~Wang, X.~Bai, C.~Gong, F.~Zhou, Hybrid inference network for few-shot sar
  automatic target recognition, IEEE Transactions on Geoscience and Remote
  Sensing (2021).

\bibitem{reduce1}
C.~Wang, J.~Shi, Y.~Zhou, X.~Yang, Z.~Zhou, S.~Wei, X.~Zhang, Semisupervised
  learning-based sar atr via self-consistent augmentation, IEEE Transactions on
  Geoscience and Remote Sensing 59~(6) (2020) 4862--4873.

\bibitem{intro6}
L.~Zhang, X.~Leng, S.~Feng, X.~Ma, K.~Ji, G.~Kuang, L.~Liu, Domain knowledge
  powered two-stream deep network for few-shot sar vehicle recognition, IEEE
  Transactions on Geoscience and Remote Sensing (2021).

\bibitem{lacking2}
Y.~Xu, H.~Lang, Ship classification in sar images with geometric transfer
  metric learning, IEEE Transactions on Geoscience and Remote Sensing 59~(8)
  (2020) 6799--6813.

\bibitem{tang2020long}
K.~Tang, J.~Huang, H.~Zhang, Long-tailed classification by keeping the good and
  removing the bad momentum causal effect, Advances in Neural Information
  Processing Systems 33 (2020) 1513--1524.

\bibitem{qin2021causal}
W.~Qin, H.~Zhang, R.~Hong, E.-P. Lim, Q.~Sun, Causal interventional training
  for image recognition, IEEE Transactions on Multimedia (2021).

\bibitem{treisman1980feature}
A.~M. Treisman, G.~Gelade, A feature-integration theory of attention, Cognitive
  psychology 12~(1) (1980) 97--136.

\bibitem{deng2019arcface}
J.~Deng, J.~Guo, N.~Xue, S.~Zafeiriou, Arcface: Additive angular margin loss
  for deep face recognition, in: Proceedings of the IEEE/CVF conference on
  computer vision and pattern recognition, 2019, pp. 4690--4699.

\bibitem{dy2004feature}
J.~G. Dy, C.~E. Brodley, Feature selection for unsupervised learning, Journal
  of machine learning research 5~(Aug) (2004) 845--889.

\bibitem{vapnik1991principles}
V.~Vapnik, Principles of risk minimization for learning theory, Advances in
  neural information processing systems 4 (1991).

\bibitem{chaudhuri2011differentially}
K.~Chaudhuri, C.~Monteleoni, A.~D. Sarwate, Differentially private empirical
  risk minimization., Journal of Machine Learning Research 12~(3) (2011).

\bibitem{cos4}
R.~Weber, H.-J. Schek, S.~Blott, A quantitative analysis and performance study
  for similarity-search methods in high-dimensional spaces, in: VLDB, Vol.~98,
  1998, pp. 194--205.

\bibitem{OpenSARShip}
L.~Zeng, Q.~Zhu, D.~Lu, T.~Zhang, H.~Wang, J.~Yin, J.~Yang, Dual-polarized sar
  ship grained classification based on cnn with hybrid channel feature loss,
  IEEE Geoscience and Remote Sensing Letters (2021).

\bibitem{FUSAR}
X.~Hou, W.~Ao, Q.~Song, J.~Lai, H.~Wang, F.~Xu, Fusar-ship: Building a
  high-resolution sar-ais matchup dataset of gaofen-3 for ship detection and
  recognition, Science China Information Sciences 63~(4) (2020) 1--19.

\bibitem{intro_aug_consit2}
L.~Huang, B.~Liu, B.~Li, W.~Guo, W.~Yu, Z.~Zhang, W.~Yu, Opensarship: A dataset
  dedicated to sentinel-1 ship interpretation, IEEE Journal of Selected Topics
  in Applied Earth Observations and Remote Sensing 11~(1) (2017) 195--208.

\bibitem{open3}
J.~He, Y.~Wang, H.~Liu, Ship classification in medium-resolution sar images via
  densely connected triplet cnns integrating fisher discrimination regularized
  metric learning, IEEE Transactions on Geoscience and Remote Sensing 59~(4)
  (2020) 3022--3039.

\bibitem{compared2}
Y.~Li, X.~Li, Q.~Sun, Q.~Dong, Sar image classification using cnn embeddings
  and metric learning, IEEE Geoscience and Remote Sensing Letters 19 (2020)
  1--5.

\bibitem{compared3}
T.~Zhang, X.~Zhang, A polarization fusion network with geometric feature
  embedding for sar ship classification, Pattern Recognition 123 (2022) 108365.

\bibitem{compared4}
H.~Zheng, Z.~Hu, J.~Liu, Y.~Huang, M.~Zheng, Metaboost: A novel heterogeneous
  dcnns ensemble network with two-stage filtration for sar ship classification,
  IEEE Geoscience and Remote Sensing Letters (2022).

\bibitem{addopensarcp3}
Y.~Zhang, J.~Xia, X.~Gao, L.~Xue, X.~Zhang, X.~Li, Sm-cnn: Separability measure
  based cnn for sar target recognition, IEEE Geoscience and Remote Sensing
  Letters (2023).

\bibitem{addopensarcp4}
Y.~Zhang, X.~Guo, L.~Li, N.~Ansari, Deep knowledge integration of heterogeneous
  features for domain adaptive sar target recognition, Pattern Recognition 126
  (2022) 108590.

\bibitem{2compared1}
Y.~LeCun, L.~Bottou, Y.~Bengio, P.~Haffner, Gradient-based learning applied to
  document recognition, Proceedings of the IEEE 86~(11) (1998) 2278--2324.

\bibitem{2compared2}
A.~Krizhevsky, I.~Sutskever, G.~E. Hinton, Imagenet classification with deep
  convolutional neural networks, Advances in neural information processing
  systems 25 (2012).

\bibitem{2compared3}
K.~Simonyan, A.~Zisserman, Very deep convolutional networks for large-scale
  image recognition, arXiv preprint arXiv:1409.1556 (2014).

\bibitem{2compared5}
C.~Szegedy, W.~Liu, Y.~Jia, P.~Sermanet, S.~Reed, D.~Anguelov, D.~Erhan,
  V.~Vanhoucke, A.~Rabinovich, Going deeper with convolutions, in: Proceedings
  of the IEEE conference on computer vision and pattern recognition, 2015, pp.
  1--9.

\bibitem{2compared6}
K.~He, X.~Zhang, S.~Ren, J.~Sun, Deep residual learning for image recognition,
  in: Proceedings of the IEEE conference on computer vision and pattern
  recognition, 2016, pp. 770--778.

\bibitem{2compared10}
G.~Huang, Z.~Liu, L.~Van Der~Maaten, K.~Q. Weinberger, Densely connected
  convolutional networks, in: Proceedings of the IEEE conference on computer
  vision and pattern recognition, 2017, pp. 4700--4708.

\bibitem{2compared12}
A.~Howard, M.~Sandler, G.~Chu, L.-C. Chen, B.~Chen, M.~Tan, W.~Wang, Y.~Zhu,
  R.~Pang, V.~Vasudevan, et~al., Searching for mobilenetv3, in: Proceedings of
  the IEEE/CVF international conference on computer vision, 2019, pp.
  1314--1324.

\bibitem{2compared13}
F.~N. Iandola, S.~Han, M.~W. Moskewicz, K.~Ashraf, W.~J. Dally, K.~Keutzer,
  Squeezenet: Alexnet-level accuracy with 50x fewer parameters and< 0.5 mb
  model size, arXiv preprint arXiv:1602.07360 (2016).

\bibitem{2compared14}
C.~Szegedy, S.~Ioffe, V.~Vanhoucke, A.~A. Alemi, Inception-v4, inception-resnet
  and the impact of residual connections on learning, in: Thirty-first AAAI
  conference on artificial intelligence, 2017.

\bibitem{2compared15}
F.~Chollet, Xception: Deep learning with depthwise separable convolutions, in:
  Proceedings of the IEEE conference on computer vision and pattern
  recognition, 2017, pp. 1251--1258.

\bibitem{p4}
Y.~Wang, C.~Wang, H.~Zhang,
  \href{https://www.mdpi.com/1424-8220/18/9/2929}{Ship classification in
  high-resolution sar images using deep learning of small datasets}, Sensors
  18~(9) (2018).
\newblock \href {https://doi.org/10.3390/s18092929}
  {\path{doi:10.3390/s18092929}}.
\newline\urlprefix\url{https://www.mdpi.com/1424-8220/18/9/2929}

\bibitem{2compared18}
G.~Huang, X.~Liu, J.~Hui, Z.~Wang, Z.~Zhang, A novel group squeeze excitation
  sparsely connected convolutional networks for sar target classification,
  International Journal of Remote Sensing 40~(11) (2019) 4346--4360.

\bibitem{2compared21}
G.~Xiong, Y.~Xi, D.~Chen, W.~Yu, Dual-polarization sar ship target recognition
  based on mini hourglass region extraction and dual-channel efficient fusion
  network, IEEE Access 9 (2021) 29078--29089.

\bibitem{2compared22}
T.~Zhang, X.~Zhang, Squeeze-and-excitation laplacian pyramid network with
  dual-polarization feature fusion for ship classification in sar images, IEEE
  Geoscience and Remote Sensing Letters 19 (2021) 1--5.

\bibitem{comparison1}
C.~Zheng, X.~Jiang, X.~Liu, Semi-supervised sar atr via multi-discriminator
  generative adversarial network, IEEE Sensors Journal 19~(17) (2019)
  7525--7533.

\bibitem{addc1}
F.~Wang, C.~Zhang, Label propagation through linear neighborhoods, in:
  Proceedings of the 23rd international conference on Machine learning, 2006,
  pp. 985--992.

\bibitem{addc2}
C.~Persello, L.~Bruzzone, Active and semisupervised learning for the
  classification of remote sensing images, IEEE Transactions on Geoscience and
  Remote Sensing 52~(11) (2014) 6937--6956.

\bibitem{addc3}
C.~Li, T.~Xu, J.~Zhu, B.~Zhang, Triple generative adversarial nets, Advances in
  neural information processing systems 30 (2017).

\bibitem{addc4}
T.~Salimans, I.~Goodfellow, W.~Zaremba, V.~Cheung, A.~Radford, X.~Chen,
  Improved techniques for training gans, Advances in neural information
  processing systems 29 (2016).

\bibitem{addc5}
F.~Gao, Y.~Yang, J.~Wang, J.~Sun, E.~Yang, H.~Zhou, A deep convolutional
  generative adversarial networks (dcgans)-based semi-supervised method for
  object recognition in synthetic aperture radar (sar) images, Remote Sensing
  10~(6) (2018) 846.

\bibitem{comparison2}
D.~A. Morgan, Deep convolutional neural networks for atr from sar imagery, in:
  Algorithms for Synthetic Aperture Radar Imagery XXII, Vol. 9475,
  International Society for Optics and Photonics, 2015, p. 94750F.

\bibitem{ac2}
Q.~Wang, H.~Xu, L.~Yuan, X.~Wen, Dense capsule network for sar automatic target
  recognition with limited data, Remote Sensing Letters 13~(6) (2022) 533--543.

\bibitem{ac3}
P.~Lang, X.~Fu, C.~Feng, J.~Dong, R.~Qin, M.~Martorella, Lw-cmdanet: A novel
  attention network for sar automatic target recognition, IEEE Journal of
  Selected Topics in Applied Earth Observations and Remote Sensing 15 (2022)
  6615--6630.

\bibitem{ac5}
C.~Zhang, H.~Dong, B.~Deng, Improving pre-training and fine-tuning for few-shot
  sar automatic target recognition, Remote Sensing 15~(6) (2023) 1709.

\bibitem{ac6}
C.~Wang, H.~Gu, W.~Su, Sar image classification using contrastive learning and
  pseudo-labels with limited data, IEEE Geoscience and Remote Sensing Letters
  19 (2021) 1--5.

\bibitem{ac9}
C.~Wang, J.~Pei, J.~Yang, X.~Liu, Y.~Huang, D.~Mao, Recognition in label and
  discrimination in feature: A hierarchically designed lightweight method for
  limited data in sar atr, IEEE Transactions on Geoscience and Remote Sensing
  60 (2022) 1--13.

\bibitem{ac10}
S.~Feng, K.~Ji, F.~Wang, L.~Zhang, X.~Ma, G.~Kuang, Electromagnetic scattering
  feature (esf) module embedded network based on asc model for robust and
  interpretable sar atr, IEEE Transactions on Geoscience and Remote Sensing 60
  (2022) 1--15.

\bibitem{fsldata4}
F.~Zhang, Y.~Wang, J.~Ni, Y.~Zhou, W.~Hu, Sar target small sample recognition
  based on cnn cascaded features and adaboost rotation forest, IEEE Geoscience
  and Remote Sensing Letters 17~(6) (2020) 1008--1012.
\newblock \href {https://doi.org/10.1109/LGRS.2019.2939156}
  {\path{doi:10.1109/LGRS.2019.2939156}}.

\bibitem{fsldata3}
Y.~Sun, Y.~Wang, H.~Liu, N.~Wang, J.~Wang, Sar target recognition with limited
  training data based on angular rotation generative network, IEEE Geoscience
  and Remote Sensing Letters 17~(11) (2020) 1928--1932.
\newblock \href {https://doi.org/10.1109/LGRS.2019.2958379}
  {\path{doi:10.1109/LGRS.2019.2958379}}.

\bibitem{readd1}
W.~Zhang, Y.~Zhu, Q.~Fu, Semi-supervised deep transfer learning-based on
  adversarial feature learning for label limited sar target recognition, IEEE
  Access 7 (2019) 152412--152420.
\newblock \href {https://doi.org/10.1109/ACCESS.2019.2948404}
  {\path{doi:10.1109/ACCESS.2019.2948404}}.

\bibitem{readd2}
{Zhang, Wei and Zhu, Yongfeng and Fu, Qiang}, Deep transfer learning based on
  generative adversarial networks for sar target recognition with label
  limitation, in: 2019 IEEE International Conference on Signal, Information and
  Data Processing (ICSIDP), IEEE, 2019, pp. 1--5.

\bibitem{ramachandra2018deep}
V.~Ramachandra, Deep learning for causal inference, arXiv preprint
  arXiv:1803.00149 (2018).

\bibitem{luo2020causal}
Y.~Luo, J.~Peng, J.~Ma, When causal inference meets deep learning, Nature
  Machine Intelligence 2~(8) (2020) 426--427.

\bibitem{cui2020causal}
P.~Cui, Z.~Shen, S.~Li, L.~Yao, Y.~Li, Z.~Chu, J.~Gao, Causal inference meets
  machine learning, in: Proceedings of the 26th ACM SIGKDD International
  Conference on Knowledge Discovery \& Data Mining, 2020, pp. 3527--3528.

\bibitem{wang2021desiderata}
Y.~Wang, M.~I. Jordan, Desiderata for representation learning: A causal
  perspective, arXiv preprint arXiv:2109.03795 (2021).

\bibitem{mitrovic2020representation}
J.~Mitrovic, B.~McWilliams, J.~Walker, L.~Buesing, C.~Blundell, Representation
  learning via invariant causal mechanisms, arXiv preprint arXiv:2010.07922
  (2020).

\bibitem{scholkopf2021toward}
B.~Sch{\"o}lkopf, F.~Locatello, S.~Bauer, N.~R. Ke, N.~Kalchbrenner, A.~Goyal,
  Y.~Bengio, Toward causal representation learning, Proceedings of the IEEE
  109~(5) (2021) 612--634.

\bibitem{oh2022learn}
K.~Oh, J.~S. Yoon, H.-I. Suk, Learn-explain-reinforce: Counterfactual reasoning
  and its guidance to reinforce an alzheimer's disease diagnosis model, IEEE
  Transactions on Pattern Analysis and Machine Intelligence (2022).

\bibitem{liu2019generative}
S.~Liu, B.~Kailkhura, D.~Loveland, Y.~Han, Generative counterfactual
  introspection for explainable deep learning, in: 2019 IEEE Global Conference
  on Signal and Information Processing (GlobalSIP), IEEE, 2019, pp. 1--5.

\bibitem{pfohl2019counterfactual}
S.~R. Pfohl, T.~Duan, D.~Y. Ding, N.~H. Shah, Counterfactual reasoning for fair
  clinical risk prediction, in: Machine Learning for Healthcare Conference,
  PMLR, 2019, pp. 325--358.

\bibitem{zhu2019causal}
S.~Zhu, I.~Ng, Z.~Chen, Causal discovery with reinforcement learning, arXiv
  preprint arXiv:1906.04477 (2019).

\bibitem{madumal2020explainable}
P.~Madumal, T.~Miller, L.~Sonenberg, F.~Vetere, Explainable reinforcement
  learning through a causal lens, in: Proceedings of the AAAI conference on
  artificial intelligence, Vol.~34, 2020, pp. 2493--2500.

\bibitem{dasgupta2019causal}
I.~Dasgupta, J.~Wang, S.~Chiappa, J.~Mitrovic, P.~Ortega, D.~Raposo, E.~Hughes,
  P.~Battaglia, M.~Botvinick, Z.~Kurth-Nelson, Causal reasoning from
  meta-reinforcement learning, arXiv preprint arXiv:1901.08162 (2019).

\end{thebibliography}

\end{document}